\documentclass{aa}  
\usepackage{graphicx, url}
\usepackage{subfigure}
\usepackage{adjustbox}
\usepackage{color}

\usepackage{amsmath}

\usepackage{natbib}

\bibpunct{(}{)}{;}{a}{}{,} 
\usepackage{txfonts}
\usepackage{hyperref} 

\hypersetup{
    colorlinks=true,
    linkcolor=blue,
    citecolor=blue,
    filecolor=magenta,      
    urlcolor=blue,
    pdftitle={Overleaf Example},
    pdfpagemode=FullScreen,
    }
\usepackage[english]{babel}

\begin{document}

   \title{Constraining the Hubble constant with scattering in host galaxies of fast radio bursts}

   \subtitle{}

   \author{Tsung-Ching Yang
          \inst{1}\fnmsep\thanks{pipi4017der@gmail.com}
          \and
          Tetsuya Hashimoto
          \inst{1}
          \and
          Tzu-Yin Hsu
          \inst{2,3}
          \and
          Tomotsugu Goto
          \inst{2,4}
          \and
          Chih-Teng Ling
          \inst{4}
          \and
          Simon C.-C. Ho\inst{5,6,7,8}
          \and
          `Amos Y.-A. Chen\inst{2}
          \and
          Ece Kilerci\inst{9}
          }

   \institute{Department of Physics, National Chung Hsing University, 145, Xingda Road, Taichung, 40227, Taiwan (R.O.C.)\\
         \and
             Department of Physics, National Tsing Hua University, 101, Section 2. Kuang-Fu Road, Hsinchu, 30013, Taiwan (R.O.C.)\\
        \and
            Institute of Astronomy and Astrophysics, Academia Sinica, 11F of AS/NTU Astronomy-Mathematics Building, No.1, Section 4, Roosevelt Road, Taipei 10617, Taiwan (R.O.C.)\\
        \and
            Institute of Astronomy, National Tsing Hua University, 101, Section 2. Kuang-Fu Road, Hsinchu, 30013, Taiwan (R.O.C.)\\
        \and
            Research School of Astronomy and Astrophysics, The Australian National University, Canberra, ACT 2611, Australia\\
        \and
            Centre for Astrophysics and Supercomputing, Swinburne University of Technology, P.O. Box 218, Hawthorn, VIC 3122, Australia\\
        \and
            OzGrav: The Australian Research Council Centre of Excellence for Gravitational Wave Discovery, Hawthorn, VIC 3122, Australia\\
        \and
            ASTRO3D: The Australian Research Council Centre of Excellence for All-sky Astrophysics in 3D, ACT 2611, Australia\\
        \and
            Sabanc{\i} University, Faculty of Engineering and Natural Sciences, 34956, Istanbul, Turkey
             }

   \date{Received 23 May 2024; accepted 21 October 2024}

 
  \abstract
   {}
   {Measuring the Hubble constant (${\rm H}_{\rm 0}$) is one of the most important missions in astronomy. 
   Nevertheless, recent studies exhibit differences between the employed methods.
   }
   {Fast radio bursts (FRBs) are coherent radio transients with large dispersion measures (DM) with a duration of milliseconds. DM$_{\rm IGM}$, the free electron column density along a line of sight in the intergalactic medium (IGM), could open a new avenue for probing ${\rm H}_{\rm 0}$.
    However, it has been challenging to separate DM contributions from different components (i.e., the IGM and the host galaxy plasma), and this hampers the accurate measurements of DM$_{\rm IGM}$ and hence ${\rm H}_{\rm 0}$. 
    We adopted a method to overcome this problem by using the temporal scattering of the FRB pulses due to the propagation effect through the host galaxy plasma (scattering time). 
    The scattering-inferred DM in a host galaxy improves the estimate of DM$_{\rm IGM}$, which in turn leads to a better constraint on ${\rm H}_{\rm 0}$.
    In previous studies, a certain value or distribution has conventionally been assumed of the dispersion measure in host galaxies (DM$_{\rm h}$).
    We compared this method with ours by generating 100 mock FRBs, and we found that our method reduces the systematic (statistical) error of H$_0$ by 9.1$\%$ (1$\%$) compared to the previous method.}
   {We applied our method to 30 localized FRB sources with both scattering and spectroscopic redshift measurements to constrain H$_0$. Our result is H$_0$ = 74$_{-7.2}^{+7.5}$ km s$^{-1}$ Mpc$^{-1}$, 
   where the central value prefers the value obtained from local measurements over the cosmic microwave background.
   We also measured DM$_{\rm h}$ with a median value of $103^{+68}_{-48}$ pc cm$^{-3}$.} 
   {The ${\rm DM_h}$ had to be assumed in previous works to derive DM$_{\rm IGM}$. 
   Scattering enables us to measure ${\rm DM_{IGM}}$ without assuming DM$_{\rm h}$ to constrain H$_0$. 
   The reduction in systematic error is comparable to the Hubble tension ($\sim10\%$). Combined with the fact that more localized FRBs will become available, our result indicates that our method can be used to address the Hubble tension using future FRB samples.}


   \maketitle
%

\section{Introduction}

    The expansion rate of the Universe is one of the most fundamental physical parameters in astrophysics. 
    The Hubble constant, ${\rm H}_{\rm 0}$, describes the relative expansion rate of the Universe. 
    ${\rm H}_{\rm 0}$ has been measured so far with different methods, such as the cosmic microwave background (CMB) \citep[e.g.,][]{refId0} and local distance ladders \citep[e.g.,][]{Riess_2022}. 
    However, there is a difference of 4 to 6$\sigma$  between the two methods, namely the CMB and the local distance ladder \citep[e.g.,][]{Verde2019,Valentino2021,Riess2021,hu2023hubble}. The recent estimate of H$_0$ from the CMB by the Planck Collaboration is 67.4 $\pm$ 0.5 km s$^{-1}$ Mpc$^{-1}$ \citep{refId0}, while the local distance ladder method by the Supernova H$_0$ for the Equation of State (SH0ES) team yielded H$_0$ = 73.0 $\pm$ 1.0 km s$^{-1}$ Mpc$^{-1}$ \citep{Riess_2022}. One possible solution to the Hubble tension might be so-called early dark energy, which introduces an additional energy density to the early Universe \citep[e.g.,][]{Hill2018}, although the difference might be explained by observational systematics \citep[e.g.,][]{Mortsell2022}.
    
    Fast radio bursts (FRBs) are enigmatic coherent radio flashes that occur at cosmological distances \citep[$z\gtrsim$0.05; e.g.,][]{Lorimer2024,Bailes2022,Petroff_2019}. They are characterized by a brief duration of approximately 1 ms and their exceptional brightness \citep[e.g.,][]{2007Sci...318..777L}.
    The dispersion measure (DM) is a unique observable of FRBs. It represents a free electron density along the line of sight to an FRB. 
    The definition of the DM is DM $\equiv \int n_e ds$, where $n_e$ is the electron number density, and $ds$ is the distance segment along the line of sight. 
    The DM is proportional to the amount of plasma along a line of sight between an FRB source and an observer. 
    Therefore, it can be used as an indicator of the redshift and distance to the FRB \citep{2007Sci...318..777L}. 
    The FRBs are critically important for addressing the issues of missing baryons \citep[e.g.,][]{Macquart_2020} and the equation of state of dark energy \citep[e.g.,][]{Zhou2014}. This made them a significant focus for further research. We rather focus on the Hubble tension in this particular paper because the existing difference between well-established methods (i.e., $\sim$10$\%$ systematics), including the CMB and the local distance ladder, underscores the importance of adding a new independent method to the comparison. The FRB method would be useful if it could achieve an accuracy better than $\sim$10$\%$ by mitigating the potential systematics in the method in the sense that it can distinguish between the H$_0$ values derived from the CMB and local distance ladder methods. Our primary focus is on leveraging the DM of FRBs as a unique distance indicator to derive H$_0$. 

    The observed DM (DM$_{\rm obs}$) can be separated into three main components. One component is DM$_{\rm MW}$, which is the DM contributed by Milky Way interstellar medium (ISM) and halo. Another is DM$_{\rm h}$, which is the DM in a galaxy hosting one FRB source (FRB host galaxy). The other is DM$_{\rm IGM}$, which is the DM in the intergalactic medium (IGM) between the Milky Way and a host galaxy,
    \begin{equation}
        \rm DM_{\rm obs} = \rm DM_{\rm MW} + \rm DM_{\rm IGM} + \frac{\rm DM_{\rm h}}{(1 + {\it z}_{\rm spec})},
    \label{DMeq}
    \end{equation}
    where $z_{\rm spec}$ is the spectroscopic redshift of the host galaxy. 
    The DM$_{\rm h}$ is divided by $1 + z_{\rm spec}$ to convert it into the observer's frame, and the DM is given in units of pc cm$^{-3}$.
    The DM$_{\rm MW}$ can be divided into two components, the disk and the halo (DM$_{\rm MW_{\rm disk}}$ and DM$_{\rm MW_{\rm halo}}$, respectively),
    \begin{equation}
        \rm DM_{\rm MW} = \rm DM_{\rm MW_{\rm disk}} + \rm DM_{\rm MW_{\rm halo}}.
    \end{equation}
    According to Eq. 9 in \citet{Zhou2014}, the cosmic average of DM$_{\rm IGM}$ is proportional to H$_0$. 
    The main focus of \citet{Zhou2014} was to discuss a constraint on the $w$ parameter for the equation of state of dark energy with FRBs. In this work, we implement their formalization (their Eq. 9) in our analysis, combined with scattering, to measure H$_0$.
    Therefore, ${\rm H}_{\rm 0}$ can be constrained with DM$_{\rm IGM}$. 
    To derive DM$_{\rm IGM}$, DM$_{\rm h}$ has to be subtracted from DM$_{\rm obs}$.
    However, in previous works \citep[e.g.,][]{Hagstotz_2022,2022arXiv221213433Z}, a certain distribution of DM$_{\rm h}$ was assumed for all FRB samples to calculate DM$_{\rm IGM}$ (see Sect. \ref{Compare paper} for details). Therefore, there might be unknown systematic uncertainties in DM$_{\rm h}$ and DM$_{\rm IGM}$ in previous works.
     
    The scattering time ($\tau$) is the pulse-broadening effect of radio pulses, including pulsars and FRBs, due to the propagation through the plasma. 
    The scattering time becomes longer when a radio pulse propagates in a larger amount of plasma with turbulence, and the scattering tail \citep{Petroff_2019} becomes more significant. 
    The scattering time has a minimal impact on the pulse-broadening effect in the Milky May and IGM \citep[e.g.,][]{Cordes_2022}. 
    Therefore, we assumed that scattering only occurs within a host galaxy. 
    The scattering time is a similar quantity as the DM$_{\rm h}$ in the sense that both increase with increasing amount of plasma in a host galaxy. 
    Therefore, the scattering time has information on DM$_{\rm h}$ and is proportional to the square of DM$_{\rm h}$ \citep{Cordes_2022}. 
    The DM$_{\rm h}$ can be measured based on the observed scattering.
    This approach is free of the potential systematics involved in the assumption on DM$_{\rm h}$, that is, either a fixed value of DM$_{\rm h}$ or a fixed shape of the distribution. 
    This marks a novel application of the scattering time in determining DM$_{\rm h}$, and it might improve the method employed to measure H$_0$.
    We focus on the capability of this approach using scattering to constrain H$_{\rm 0}$, while \citet{Cordes_2022} used it to constrain the fraction of baryons in the IGM.
    
    Throughout this paper, we assume the Planck 2018 results implemented in Astropy \citep{refId0}, that is, a $\Lambda$ cold dark matter cosmology with ($\Omega_{\rm m}$, $\Omega_{\rm b}$) $=$ (0.31, 0.049).

\section{Method}
\label{S:data}
We provide an overview of the method for constraining H$_{\rm 0}$ with DM and scattering.
    We followed the formalization presented in \citet{Cordes_2022}.
    \citet{Cordes_2022} constrained the fraction of baryons in the intergalactic space with a given H$_{0}$.
    We instead focus on how accurately H$_{0}$ can be constrained under their formalization.
    The scattering time ($\tau$) was used to measure a probability density function (PDF) of ${\rm DM_h}$, where the PDF is described as a function of two parameters: DM$_{\rm h}$ and $A_{\tau}\tilde{F} G$.
    These two parameters are discussed in detail in Sect. \ref{SE:Photometry}.
    The integration of the PDF over all possible ranges of the two parameters \citep{Cordes_2022} was computed as a function of the redshift to derive the PDF of redshift.
    Using the 50th percentile of the PDF of the redshift, we can then determine a modeled redshift ($z_{\rm model}$) that is derived by using DM and $\tau$. 
    $z_{\rm model}$ can be optimized to the observed redshift ($z_{\rm spec}$) by changing ${\rm H}_{\rm 0}$.
    The best-fit $z_{\rm model}$ provides us with the measurement of ${\rm H}_{\rm 0}$.
    
    \subsection{Measuring \texorpdfstring{${\rm DM_h}$}{DMh}}
    \label{SE:Photometry}
    
    We adopted this equation to describe the rest-frame $\tau_{\rm rest}$ as a function of $\rm DM_{\rm h}$,
    \begin{equation}
        \tau_{\rm rest} ({\rm DM_{h}}, v_{\rm rest} ) = C_{\tau} v_{\rm rest}^{-4} A_{\tau}\tilde{F} G \left( \frac{\rm DM_{\rm h}}{100}\right) ^2,
    \label{yaya}
    \end{equation}
    where $v_{\rm rest}$ is the rest-frame frequency in units of GHz. 
    The quantity $C_{\tau}=0.48$ ms \citep{Cordes_2022} is a numerical constant. 
    To account for how empirical estimates for the scattering time are related to the $e^{-1}$ time, \citet{Cordes_2022} introduced a dimensionless factor $A_{\tau}$.
    $\tilde{F}$ (pc$^{2}$ km)$^{-1/3}$ is a parameter that characterizes density fluctuations.
    $G$ is a dimensionless geometric factor. 
    While the three parameters $A_{\tau}$, $\tilde{F}$, and $G$ have different physical meanings, they appear in the formalization as a product.
    Therefore, we treated the product $A_{\tau}\tilde{F} G$ as a single parameter.
    The prior assumption on the $A_{\tau}\tilde{F} G$ range is from 0.001 to 10 (pc$^2$ km)$^{-\frac{1}{3}}$ \citep{Cordes_2022}.
    
    Two parameters, ${\rm DM_h\ (pc \ cm^{-3})}$ and $A_{\tau}\tilde{F} G$ ($\phi$) ${\rm (pc^2 \ km)^{-\frac{1}{3}}}$, are used to describe the PDF of ${\rm DM_h}$ \citep{Cordes_2022}:
    \begin{equation}
    \begin{aligned}
        &f_{\rm DM_{h},\phi}({\rm DM_{h}},\phi \vert {\rm DM_{\rm obs}},z_{\rm spec},\tau_{\rm obs}) \\
        &\propto f_{\rm DM_{h}}({\rm DM_{h}}\vert {\rm DM_{\rm obs}},z_{\rm spec}) \times f_{\tau}(\tau_{\rm obs} - \hat{\tau}).
    \label{3D}
    \end{aligned}
    \end{equation}
    In Eq. \ref{3D}, there are two terms on the right side: the first term uses DM, and the second term uses $\tau$.
    The first term, $f_{\rm DM_{h}}$, is expressed as 
    \begin{equation}
    \begin{aligned}
        &f_{\rm DM_{h}}({\rm DM_{h}}\vert {\rm DM_{\rm obs},DM_{MW}},z_{\rm spec})\\&=\frac{f_{{\rm DM_{IGM}}}({\rm DM_{IGM}}\vert {\rm DM_{\rm obs}, DM_{MW}}, z_{\rm spec})} { 1+z_{\rm spec}}.
    \label{3D2}
    \end{aligned}
    \end{equation}
    $f_{\rm DM_{h}}$ is determined using the PDF of DM$_{\rm IGM}$ ($f_{DM_{\rm IGM}}$), where DM${\rm _{IGM}=DM_{obs}-DM_{MW_{disk}}-DM_{MW_{halo}}}-\frac{\rm DM_h}{1+z_{\rm spec}}$. 
    The uncertainty of DM$_{\rm MW}$ can be accounted for by introducing prior PDFs of DM$_{\rm MW_{\rm disk}}$ and DM$_{\rm MW_{\rm halo}}$ in Eq. \ref{3D2}.
    We assumed a flat PDF for DM$_{\rm MW_{disk}}$ centered at the mean value derived from the NE2001 \citep{cordes+2003} with $\pm 20\%$ deviations \citep{Cordes_2022}. 
    A flat PDF of $\rm DM_{\rm MW_{\rm halo}}$ was assumed. It ranged from 25 to 80 pc cm$^{-3}$ \citep{Prochaska&Neeleman+2018,Shull&Danforth+2018,Prochaska&Zheng+2019,Yamasaki&Totani+2020}.
    We used a log-normal distribution to characterize $f_{\rm DM_{\rm IGM}}$ \citep{Cordes_2022}. 
    The log-normal in the form, $N(\mu,\sigma)$, is a function of the following parameters:
    
    \begin{equation}
        \mu=\ln{\left( \overline{\rm DM}_{\rm IGM}\right)}-\frac{\sigma^2}{2},
    \end{equation}
    
    \begin{equation}
    \sigma = \sqrt{ \ln\left( 1 + \left( \frac{\sigma_{\rm DM_{\rm IGM}}}{\overline{\rm DM}_{\rm IGM}} \right)^2 \right) },
    \end{equation}

    with the cosmic variance of DM$_{\rm IGM}$ ($\sigma_{\rm DM_{\rm IGM}}$).
    $\sigma_{\rm DM_{\rm IGM}}$ is described as follows:
    
    \begin{equation}
        \sigma_{{\rm DM_{IGM}}(z_{\rm spec})} = \sqrt{\overline{\rm DM}_{\rm IGM}(z_{\rm spec})\  {\rm DM_c}},
    \label{DM_IGM_e}
    \end{equation}
    where DM$_{\rm c}$ is a constant, that is, DM$_{\rm c}$ = 50 pc cm$^{-3}$\citep{McQuinn_2014}, and $\overline{\rm DM}_{\rm IGM}$ is the cosmic average of DM in the nonuniform intergalactic medium.
   The  $\overline{\rm DM}_{\rm IGM}$ is described as follows:
    
    \begin{equation}
    \begin{aligned}
        &\overline{\rm DM}_{\rm IGM}\\
        &= {\rm H}_{\rm 0} \times f_{\rm IGM} \frac{3\Omega_{b} c(Y_{H}+\frac{1}{2}Y_p)}{8\pi G m_p} \int_{0}^{z}\frac{1+z_{\rm spec}}{\Omega_m (1+z_{\rm \rm spec})^{3} +1-\Omega_m}dz
    \label{gg}
    \end{aligned}
    \end{equation}
    for a flat $\Lambda$CDM Universe.
   The proton mass is $m_{p} = 1.67 \times 10^{-27}$ kg, and $Y_{H}=\frac{3}{4}$ and $Y_{p}=\frac{1}{4}$ are the mass fractions of hydrogen and helium, respectively \citep{Zhou2014}. 
  The fraction of baryons in the IGM is  $f_{\rm IGM}=0.85 \pm 0.05$ \citep{Cordes_2022}. 
   The matter and baryon density are $\Omega_{m}$ and $\Omega_{b}$ \citep{Zhou2014}, respectively.
    For the cosmological parameters, we used the Planck 2018 results implemented in Astropy \citep{refId0}.
    
    In the second term of Eq. \ref{3D}, $f_{\rm \tau}$ is the PDF of $\tau_{\rm obs}-\hat{\tau}$, where $\tau_{\rm obs}$ is the observed scattering, and $\hat{\tau}$ is the theoretical value of the scattering.
    $\hat{\tau}$ is described as follows:
    
    \begin{equation}
        \hat{\tau} = C_{\tau} v_{\rm obs}^{-4} \phi \left(\frac{\rm DM_{h}}{100}\right)^2 (1+z_{\rm spec})^{-3}, 
    \end{equation}
    where $C_{\tau}=0.48$ ms \citep{Cordes_2022}. 
   The observed frequency $v_{\rm obs}$ is given in units of GHz, and $\phi$ is the $A_{\tau} \tilde{F} G$ parameter \citep{Cordes_2022}.
    
\begin{figure}
        \centering
        \resizebox{\hsize}{!}{\includegraphics{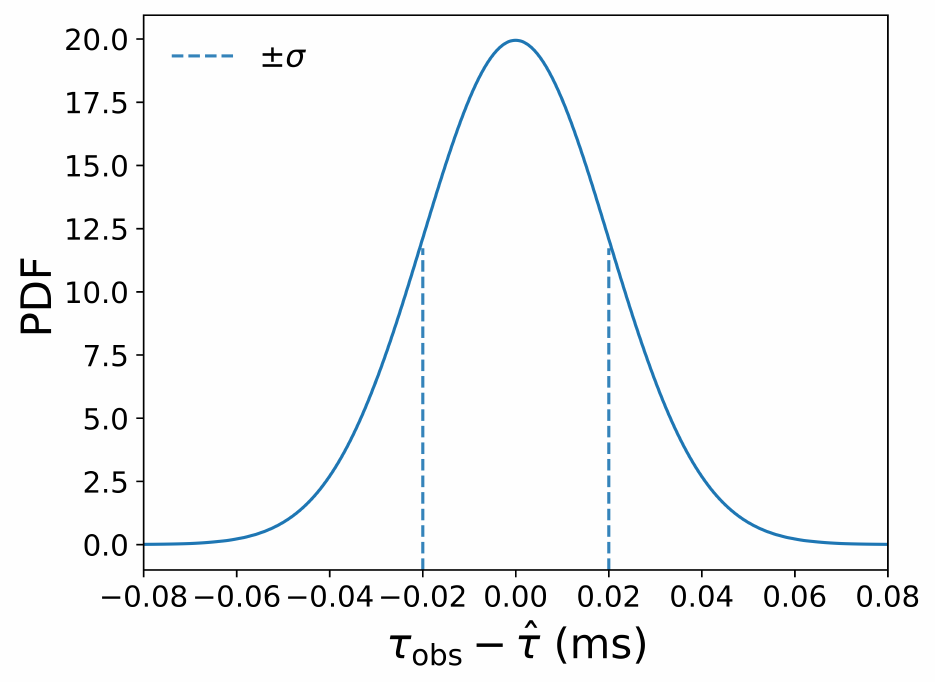}}
        \resizebox{\hsize}{!}{\includegraphics{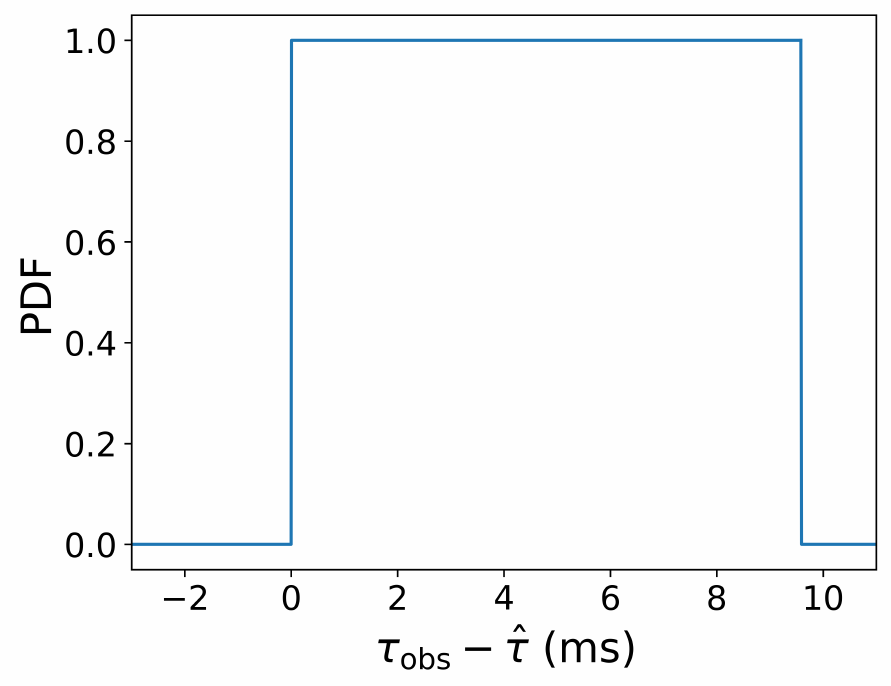}}
        
        \caption {
        PDF of $\tau_{\rm obs} - \hat{\tau}$. 
        Top panel: Case for FRB 20220509G as an example (see Sect. \ref{samples} for details), assuming $\log$ $(A_{\tau} \tilde{F} G) = -1$.
        The peak of the PDF is at $\tau_{\rm obs} - \hat{\tau}=0$, and the standard deviation is $\sigma_{\tau}= 0.02$ ms (see also Table \ref{cat}). 
        Bottom panel: FRBs with an upper limit of the scattering were given a flat PDF. 
        We take FRB 20121102A as an example with the observed scattering of $\tau_{\rm obs} <9.6$ ms (Sect. \ref{samples} and Table \ref{cat}), assuming $\log$ $(A_{\tau} \tilde{F} G) = -1$.
        When $\hat{\tau}$ is lower than 0 or greater than $\tau_{\rm obs}$, the probability is 0. 
        When $\hat{\tau}$ falls between 0 and $\tau_{\rm obs}$, the probability is nonzero and flat.
        We note that the vertical axes in both panels indicate relative quantities of the probability density. 
        }
        \label{ftau}
    \end{figure}

    A Gaussian function was assumed to express $f_{\tau}(\tau_{\rm obs} - \hat{\tau})$ (Fig. \ref{ftau} top). 
    In the Gaussian function, the mean value was 0, and the observed error in the scattering ($\sigma_{\tau}$) was adopted as the standard deviation. 
    However, some FRBs only have upper limits of the scattering. 
    In these cases, we assumed a flat PDF of $\tau_{\rm obs} - \hat{\tau}$ (Fig. \ref{ftau} bottom).
    We adopted $3\sigma_{\tau}$ as the upper bound of the flat PDF.
    
\begin{figure*}
    \centering
    \subfigure[FRB20181112A]{
        \begin{minipage}[b]{0.46\textwidth} 
        \includegraphics[width=\columnwidth]{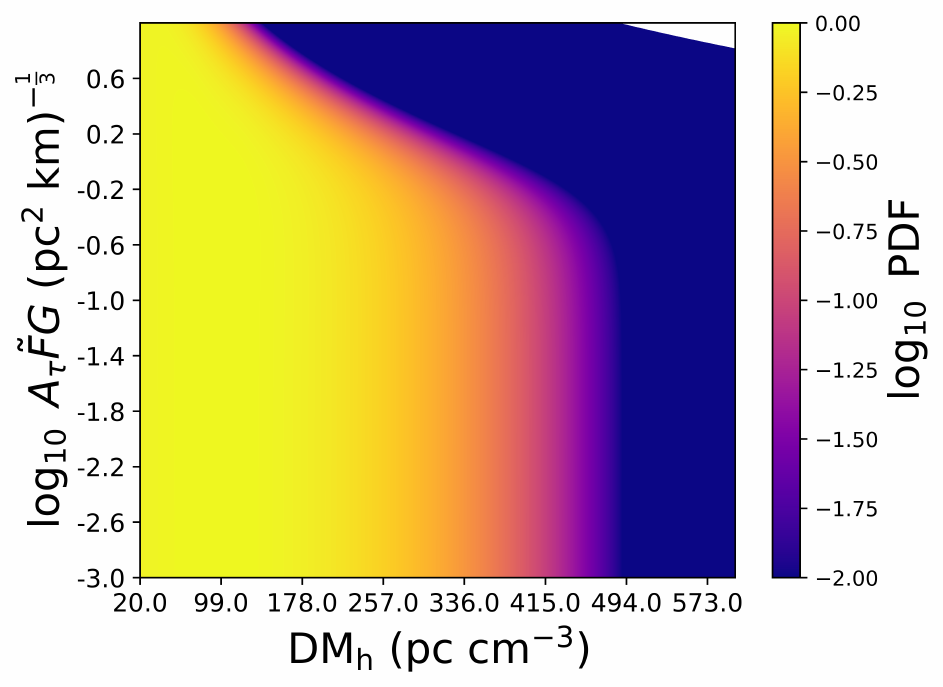}
        \includegraphics[width=\columnwidth]{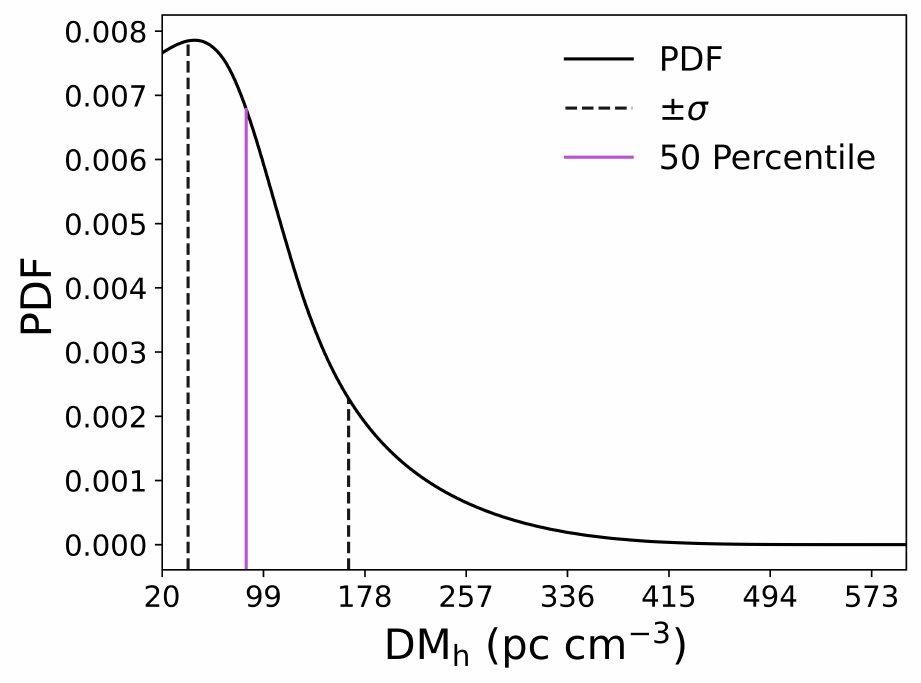}       
        \end{minipage}}
    \quad 
    \qquad 
    \subfigure[FRB20191001A]{
        \begin{minipage}[b]{0.46\textwidth} 
        \includegraphics[width=\columnwidth]{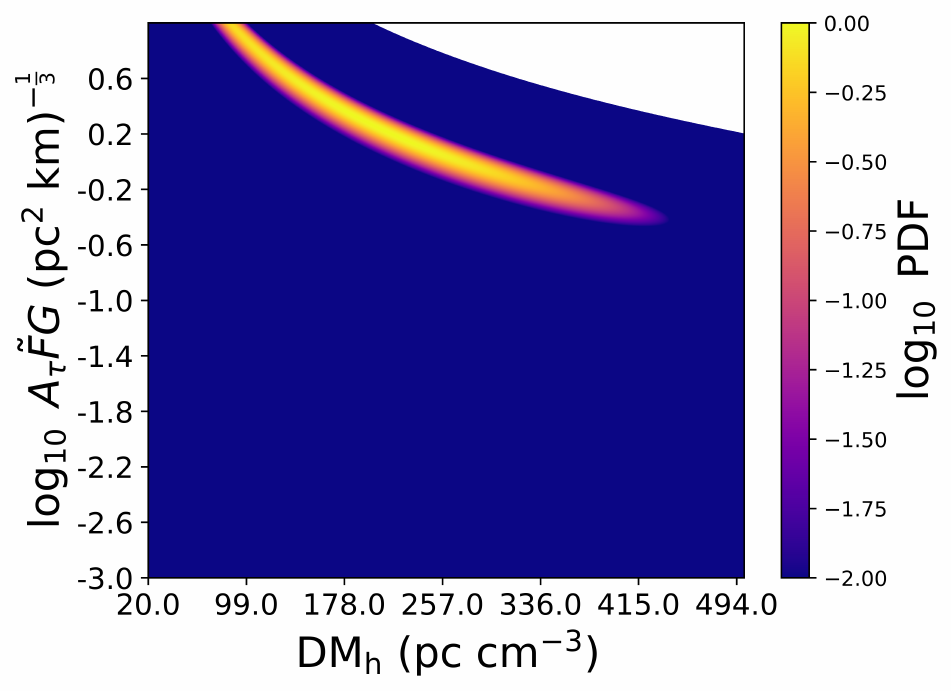}
        \includegraphics[width=\columnwidth]{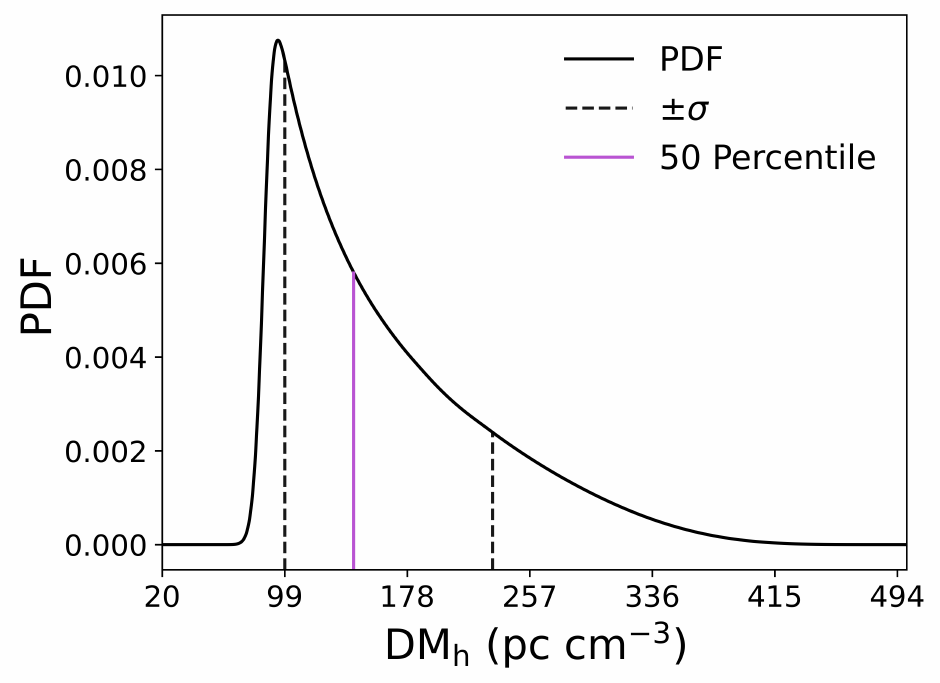}               
        \end{minipage}}
    \caption{Two examples of the PDF in our observational samples. Top left panel: Three-dimensional image of  $f_{\rm DM_{h}}$ for FRB 20181112A (see Sect. \ref{samples} and Table \ref{cat} for details). 
   The  $x$-axis is the $\rm DM_h$ parameter from 20 to 600 pc cm$^{-3}$. 
  The   $y$-axis is the $A_{\tau} \tilde{F} G$ parameter from $-3$ to 1 $(\rm pc^2 \ \rm km)^{-\frac{1}{3}}$ in log scale. 
    For a given ${\rm DM_h}$ and $A_{\tau} \tilde{F} G$, the corresponding probability density in the log scale is presented by color. 
    Bottom left panel: Integration of the PDF over the $A_{\tau} \tilde{F} G$ parameter from 0.001 to 10 $(\rm pc^2 \ \rm km)^{-\frac{1}{3}}$, which provides us with the PDF of ${\rm DM_h}$. 
    The purple line shows the 50 percentile of the PDF. 
    The dashed black lines correspond to the 84.2 and 15.8 percentiles of the PDF ($\pm \sigma$). 
    The 50 percentile is ${\rm DM_h} = 85.6$ pc cm$^{-3}$, $+\sigma = 78$ pc cm$^{-3}$, and $-\sigma = 45.4$ pc cm$^{-3}$.
    Top right panel: Example of a 3D image of $f_{\rm DM_{h}}$ by using FRB 20191001A (see Sect. \ref{samples} and Table \ref{cat} for details). 
    Bottom left panel: Same as the bottom left panel, but using FRB 20191001A. 
    The 50 percentile is ${\rm DM_h} = 143.3$ pc cm$^{-3}$, $+\sigma = 89.3$ pc cm$^{-3}$, and $-\sigma = 44.2$ pc cm$^{-3}$.}
    \label{fdmh}
    \end{figure*}

    To visualize Eq. \ref{3D}, we show the PDF of DM$_{\rm h}$ as a function of DM$_{\rm h}$ and $A_{\tau} \tilde{F} G$ in the top panels of Fig. \ref{fdmh}, where ${\rm H}_{\rm 0}=74.3$ km s$^{-1}$ Mpc$^{-1}$ was assumed. 
    By integrating $A_{\tau} \tilde{F} G$ from 0.001 to 10 $(\rm pc^2 \ \rm km)^{-\frac{1}{3}}$, we present the PDF of ${\rm DM_h}$in the bottom panels of Fig. \ref{fdmh}.
    
    
    \subsection{Redshift PDF}\label{PDF of redshift}
    We used the following equation to calculate the redshift PDF ($f_{z}$) with the parameters of DM$_{\rm obs}$ and $\tau_{\rm obs}$:
    \begin{equation}
    \begin{aligned}
        &f_{z}(z \vert {\rm DM_{obs}},\tau_{\rm obs}) \\
        &\propto \iint d{\rm DM_{h}}d\phi ~f_{\rm DM_{h},\phi}({\rm DM_{h}},\phi) \times f_{\tau}(\tau_{\rm obs} - \hat{\tau}),
    \end{aligned}
    \label{PDF_z}
    \end{equation}
    where the integration range of ${\rm DM_h}$ was $[20,1600]$ pc cm$^{-3}$ \citep{Cordes_2022}. 
    In Eq. \ref{PDF_z}, we calculate the integration of all possible ${\rm DM_h}$ and $A_{\tau} \tilde{F} G$, which provided us with $f_z$.
    \begin{figure}
        \centering
        \includegraphics[width=\columnwidth]{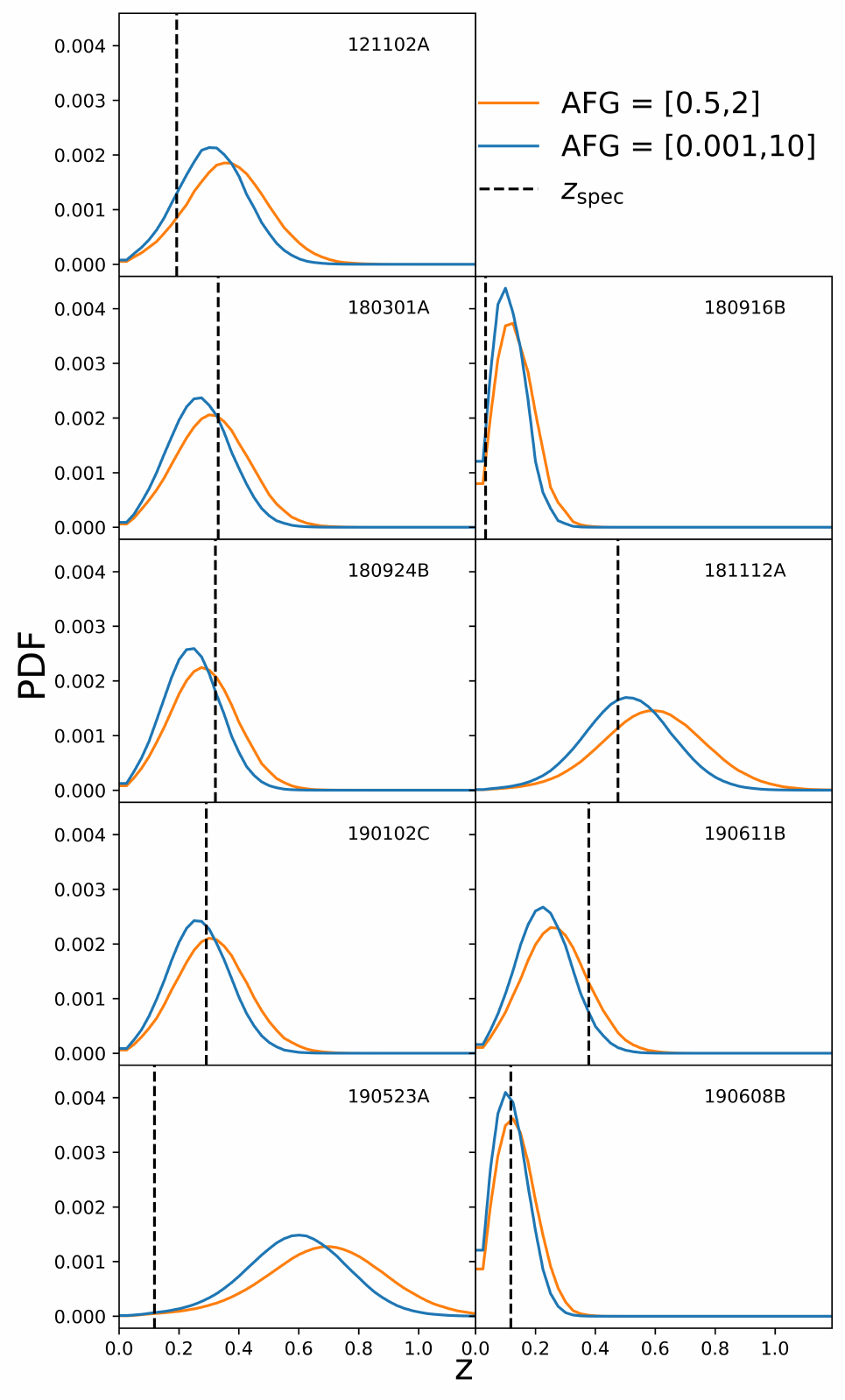}
        \caption {Nine examples representing the PDFs of redshift ($z$). 
        These were randomly selected from our FRB samples (see Sect. \ref{samples} and Table \ref{cat} for details). 
        ${\rm H}_{\rm 0} = 74.3$ km s$^{-1}$ Mpc$^{-1}$ and ${\rm DM_h}=[20,1600]$ pc cm$^{-3}$ were adopted. 
        In each panel, we present PDFs based on two different ranges of $A_{\tau} \tilde{F} G$: [0.5, 2] (orange line) and [0.001, 10] (blue line) $({\rm pc}^2 \ \rm km)^{-\frac{1}{3}}$. 
        The vertical dashed black line shows the observed redshift ($z_{\rm spec}$) in each panel.}
        \label{pdfz}
    \end{figure}
    
    To demonstrate $f_{z}$, we show examples of nine FRBs assuming ${\rm H}_{\rm 0} = 74.3$ km s$^{-1}$ Mpc$^{-1}$ in Fig. \ref{pdfz} . 
    Following \citet{Cordes_2022}, we present two prior assumptions on the $A_{\tau} \tilde{F} G$ range in this work: narrow [0.5, 2] and wide [0.001, 10] $({\rm pc}^2 \ \rm km)^{-\frac{1}{3}}$. 
    The possible impact on the H$_{0}$ measurement with different prior assumptions on the $A_{\tau} \tilde{F} G$ parameter is discussed in Sects. \ref{simulation} and \ref{sec:result}.  
    Throughout the paper, we present the results assuming the wide range of $A_{\tau} \tilde{F} G =[0.001, 10]$ $({\rm pc}^2 \ \rm km)^{-\frac{1}{3}}$ as a fiducial model unless otherwise mentioned.
    \subsection{Optimize \texorpdfstring{${\rm H}_{\rm 0}$}{H0}}
    \label{opt_H0}
\begin{figure*}
      \centering
      \includegraphics[width=17cm]{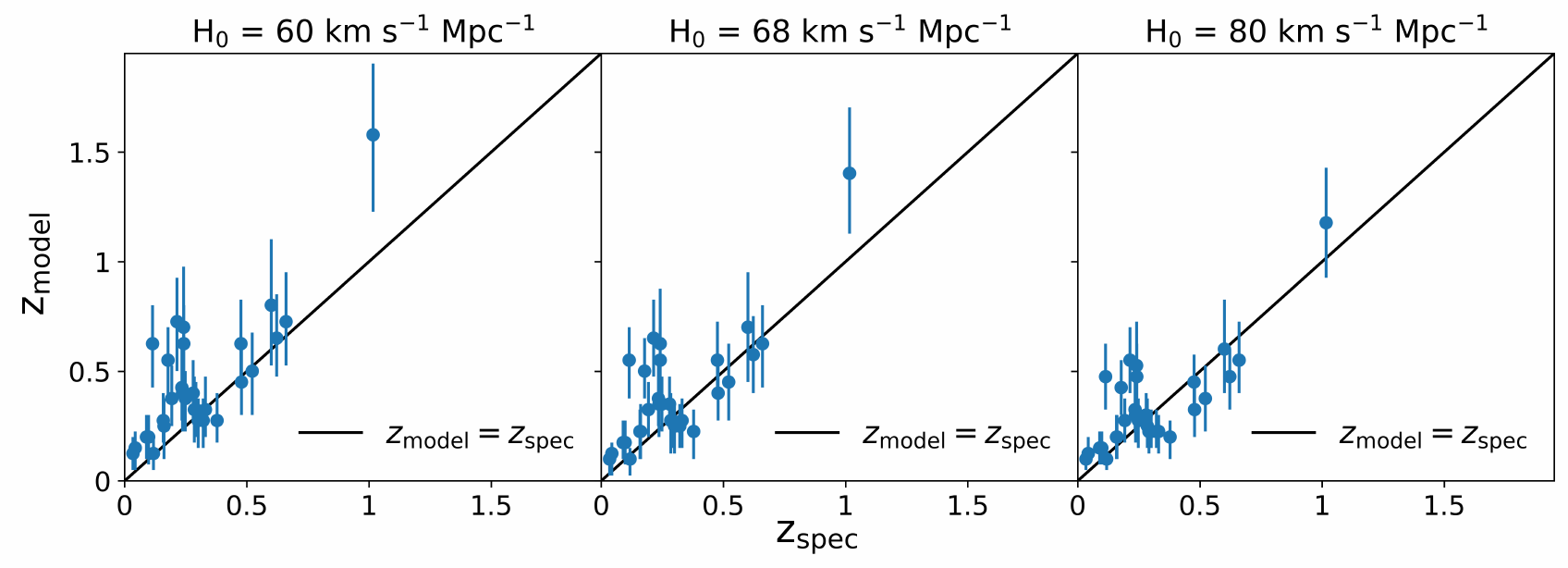}
      \caption{Observed redshift ($z_{\rm spec}$) v.s. modeled redshift ($z_{\rm model}$) to optimize ${\rm H}_{\rm 0}$. 
      To demonstrate how $z_{\rm model}$ depends on H$_{0}$, we present three values of ${\rm H}_{\rm 0}$, 60, 68, and 80 km s$^{-1}$ Mpc$^{-1}$, in the different panels. 
      We used 30 FRBs (see Sect. \ref{samples} and Table \ref{cat} for details) with ${\rm DM_h}=[20,1600]$ pc cm$^{-3}$ and $A_{\tau} \tilde{F} G = [0.001,10]$ $({\rm pc}^2 \ \rm km)^{-\frac{1}{3}}$. 
      The black line corresponds to $z_{\rm model}=z_{\rm spec}$.}
      \label{Optimize}
    \end{figure*}

    We chose the 50 percentile of $f_{z}$ to define the modeled redshift ($z_{\rm model}$). 
    By letting ${\rm H}_{\rm 0}$ be a parameter (Eq. \ref{gg}), $z_{\rm model}$ is a function of ${\rm H}_{\rm 0}$. 
    To optimize ${\rm H}_{\rm 0}$, we compared $z_{\rm model}$ and $z_{\rm spec}$, and we adjusted H$_{0}$ so that these two quantities became consistent within their errors, that is, $z_{\rm model}=z_{\rm spec}$.
    In Fig. \ref{Optimize}, we show how $z_{\rm model}$ depends on ${\rm H}_{\rm 0}$. 
    
    We only changed H$_{0}$ in Fig. \ref{Optimize} to demonstrate how this method works using 30 FRB samples (see Sect. \ref{samples} and Table \ref{cat} for details).
    However, H$_{0}$ degenerates with $f_{\rm IGM}$ in Eq. \ref{gg} because the observational error of $f_{\rm IGM}$ is $\sim$6\% \citep[e.g.,][]{Li+2019, Cordes_2022}. 
    Therefore, we present two cases of (i) $f_{\rm IGM}\times {\rm H}_{\rm 0}$ as a single parameter in Eq. (\ref{gg}) and (ii) H$_{\rm 0}$ alone for a given $f_{\rm IGM}$ and its error in Sect. \ref{sec:result}. 
    \\
\section{Simulation using mock FRB samples}
\label{simulation}
\subsection{Our method using scattering}
\label{simu_our_method}
\begin{figure}
        \centering
        \includegraphics[width=\columnwidth]{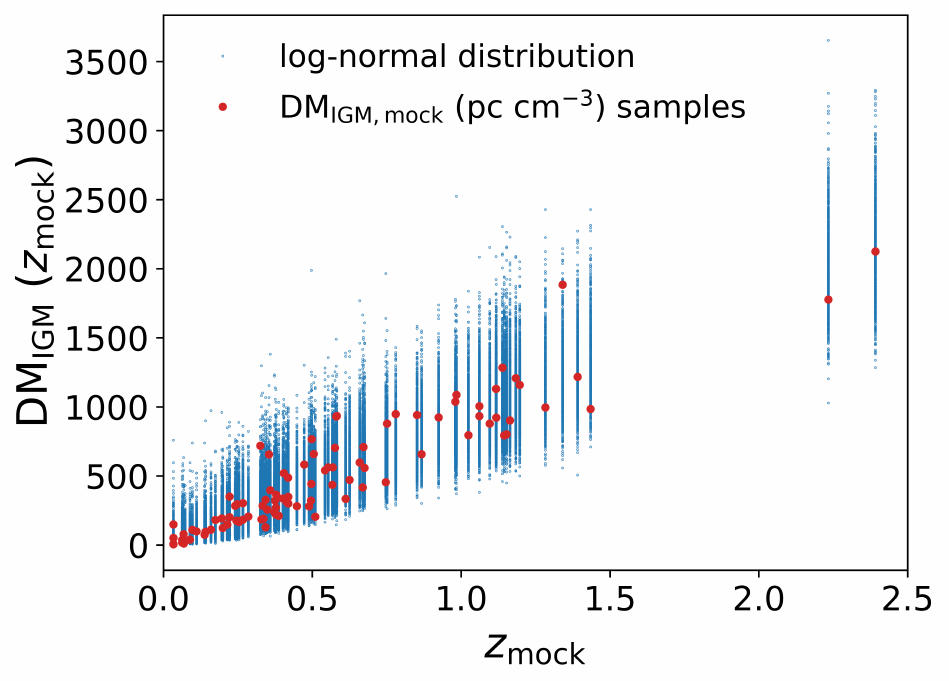}
        \caption{
        DM$_{\rm IGM, mock}$ as a function of $z_{\rm mock}$. 
        The red dots represent DM$_{\rm IGM, mock}$ generated by using Eq. \ref{DM_IGM_e} at each $z_{\rm mock}$, taking the line-of-sight fluctuation of DM$_{\rm IGM}$ into account.
        For a given $z_{\rm mock}$, we randomly selected one mock data point (red dot) from the blue dots. 
        We iterated this process 100 times at the different $z_{\rm mock}$, generating 100 mock FRB data.
        }
    \label{mock DMIGM}
    \end{figure}

    Before we applied our method to the observed data, we demonstrate how accurate our measurements are compared with a previous method via simulations.
    Our aim is a prediction based on a reasonable sample size that can be achieved in the near future. With future instruments such as the CHIME Outrigger \citep{Mena_Parra+2022} and BURSTT \citep{2022PASP..134i4106L,2023ApJ...950...53H}, it would be feasible to identify 100 FRB host galaxies. Moreover, \cite{James_2022} predicted that a sample of approximately 100 FRBs with spectroscopic redshifts would be sufficient to distinguish the $\sim$10$\%$ systematic discrepancies of the Hubble tension between the distance ladder and CMB methods. Therefore, we followed \cite{James_2022} to decide the sample size of the mock data.
    We generated 100 mock FRBs as follows. 
    First, we randomly selected redshift values for 100 mock FRBs from the CHIME FRB catalog \citep{Amiri_2021} based on the DM-derived redshift in \citet{2022MNRAS.511.1961H} to create mock redshifts, $z_{\rm mock}$. 
    Then, we randomly sampled NE2001 values for 100 FRBs from the same catalog to create mock DM$_{\rm MW_{disk}}$ values (DM$_{\rm MW_{disk, mock}}$). 
    The adopted value of the mock DM$_{\rm MW_{halo}}$ (DM$_{\rm MW_{halo, mock}}$) is 52.5 pc cm$^{-3}$. 
    Using $z_{\rm  mock}$, we calculated DM$_{\rm IGM}$ based on Eq. (\ref{gg}) assuming H$_0$ = 68 km s$^{-1}$ Mpc$^{-1}$ and $f_{\rm IGM} = 0.85$, 
    where the line-of-sight fluctuation of DM$_{\rm IGM}$ was taken into account based on Eq. (\ref{DM_IGM_e}) to create mock DM$_{\rm IGM}$ values (DM$_{\rm IGM, mock}$). 
    Figure \ref{mock DMIGM} illustrates the generation of 100 DM$_{\rm IGM, mock}$. 
    
    In general, DM$_{\rm h}$ is highly uncertain because DM$_{\rm obs}$ is an integrated quantity along a line of sight to an FRB, and hence, it is hard to separate between DM$_{\rm h}$ and DM$_{\rm IGM}$ without using scattering.
    For instance, \cite{Macquart_2020} assumed DM$_{\rm h} = 50$ pc cm$^{-3}$ in their analysis. 
    In contrast, \citet{RR_2021} conducted a cross-correlation analysis between CHIME FRB samples and galaxy catalogs to conclude that DM$_{\rm h}$ is about 400 pc cm$^{-3}$, statistically.
    An even larger average DM$_{\rm h}$ might be suggested by a population-model approach \citep[e.g.,][]{Wang2024}.
    These works highlight the significant uncertainty of the typical value of DM$_{\rm h}$, ranging from $\sim$50 pc cm$^{-3}$ to $\sim$400 pc cm$^{-3}$.
    To generate mock FRBs, we assumed a normal distribution with a mean of 200 pc cm$^{-3}$ and a standard deviation of 50 pc cm$^{-3}$ to create mock DM$_{\rm h}$ values (DM$_{\rm h, mock}$). 
    
    Subsequently, using Eq. (\ref{DMeq}), we derived 100 mock DM$_{\rm obs}$ values (DM$_{\rm obs, mock}$) from the $z_{\rm mock}$, DM$_{\rm IGM, mock}$, DM$_{\rm MW, mock}$, and DM$_{\rm h, mock}$. 
    The mock scattering times ($\tau_{\rm mock}$) at the host frame 1 GHz for the 100 mock FRBs were determined using DM$_{\rm h, mock}$ with $A_{\tau} \tilde{F} G=$ 1 $({\rm pc}^2 \ \rm km)^{-\frac{1}{3}}$ and $v=1$ GHz from Eq. \ref{yaya}.
    $\tau_{\rm mock}$ at the observer's frame ($\tau_{\rm mock,obs}$) was calculated by dividing $\tau_{\rm mock}$ by a $(1+z_{\rm mock})^{3}$ factor, assuming $\nu^{-4}$ dependence of $\tau$ and a time-dilation factor of $(1+z)$.
    The mock error in scattering ($\sigma_{\tau,\rm mock}$) was simulated based on the median value of the fractional errors of $\tau_{\rm obs}$ in Table (\ref{cat}) (i.e., $\sigma_{\tau,\rm mock}$ = 5.2$\%$). 
    In summary, we generated 100 mock FRBs including $z_{\rm mock}$, DM$_{\rm IGM, mock}$, DM$_{\rm MW, mock}$, DM$_{\rm h, mock}$, DM$_{\rm obs, mock}$, $\tau_{\rm mock,obs}$, and $\sigma_{\tau,\rm mock}$. 
    
\begin{figure}
        \centering
        \includegraphics[width=\columnwidth]{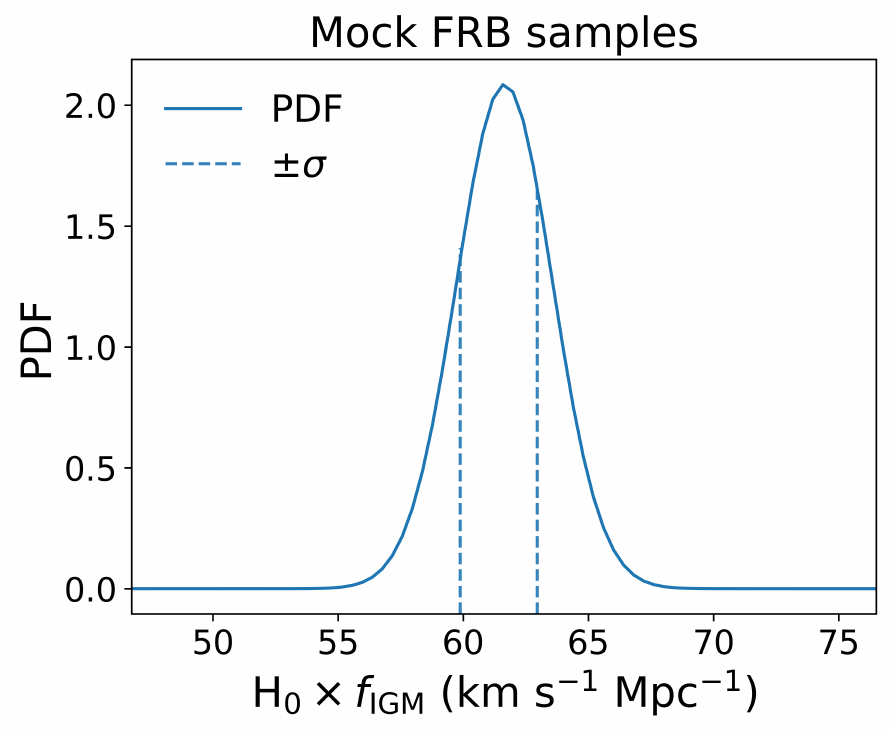}
        \includegraphics[width=\columnwidth]{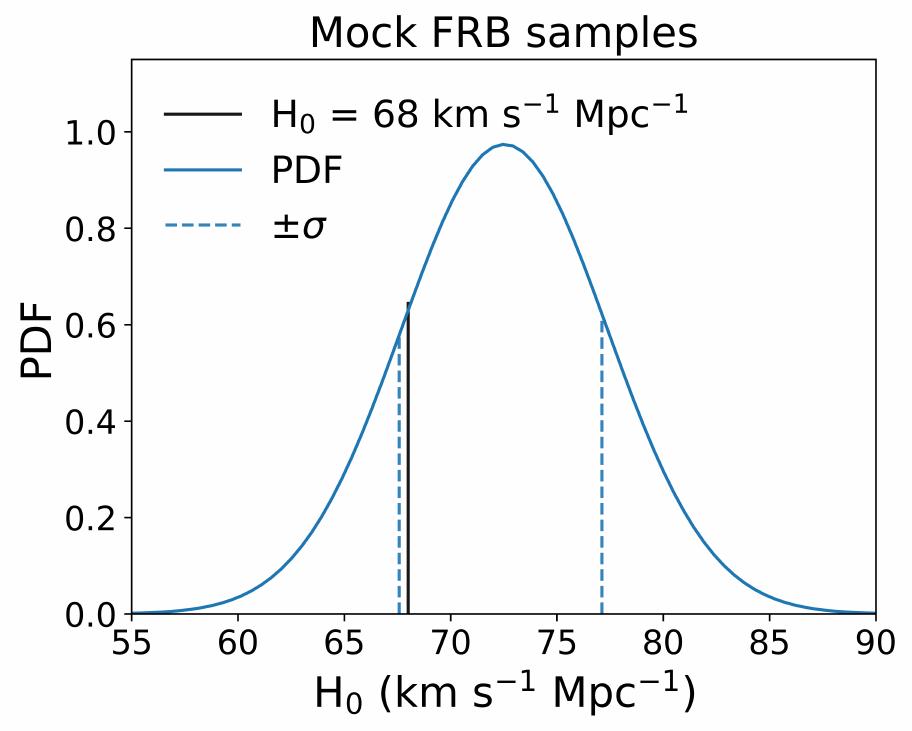}
        \caption{Simulations of our method with mock FRB samples. Top panel: PDF of H$_{\rm 0}\times f_{\rm IGM}$ derived by using the 100 mock FRBs and our method with scattering (blue curve). The vertical dashed blue lines correspond to 84.2 and 15.8 percentiles ($\pm\sigma$) of the PDF. Bottom panel: PDF of H$_0$ by changing the scale from H$_{\rm 0}\times f_{\rm IGM}$ (top panel) to H$_0$. To generate the mock data, we assumed H$_{\rm 0} = 68.0$ km s$^{-1}$ Mpc$^{-1}$, which is shown by the solid vertical black line.
        The vertical dashed blue lines are the positive and negative standard deviations of H$_0$ after taking the $f_{\rm IGM}$ error into account,
        where the $f_{\rm IGM}$ error is $\pm$0.05 \citep{Cordes_2022}.
        }
    \label{mock PDF}
    \end{figure}

    Figure \ref{mock PDF} displays the PDF of H$_{\rm 0}\times f_{\rm IGM}$ and H$_0$ derived by applying our method to the 100 mock FRBs. 
    The result is H$_0$$\times f_{\rm IGM} = 61.6^{+1.6}_{-2.0}$ km s$^{-1}$ Mpc$^{-1}$.
    This value corresponds to H$_0 = 72.5^{+4.7}_{-4.9}$ km s$^{-1}$ Mpc$^{-1}$, where we assumed $f_{\rm IGM}=0.85$ taking the $\pm 0.05$ error of $f_{\rm IGM}$ \citep{Cordes_2022} into account. 
    This reconstructed H$_{\rm 0}$ is consistent with the assumption made in the simulation, that is, H$_{\rm 0} = 68.0$ km s$^{-1}$ Mpc$^{-1}$ within the statistical error.

\subsection{Prior assumptions on the \texorpdfstring{$A_{\tau} \tilde{F} G$}{AFG} range in our method}
\label{priorAFG}
In Eq. \ref{PDF_z}, the prior assumption on the $A_{\tau} \tilde{F} G$ range needs to be specified.
Following \citet{Cordes_2022}, we briefly present how the narrow and wide $A_{\tau} \tilde{F} G$ ranges affect the H$_{\rm 0}$ measurements based on the 100 FRB mock data.
We applied the method described in Sect. \ref{S:data} to the 100 mock data, assuming $A_{\tau} \tilde{F} G =$ [0.5, 2] and [0.001, 10] $({\rm pc}^2 \ \rm km)^{-\frac{1}{3}}$.
For a given value of $f_{\rm IGM}=0.85\pm 0.05$, the derived H$_{\rm 0}$ are 70.6$^{+4.4}_{-4.8}$ and 72.5$^{+4.7}_{-4.9}$ km s$^{-1}$ Mpc$^{-1}$ for $A_{\tau} \tilde{F} G =$ [0.5, 2] and [0.001, 10] $({\rm pc}^2 \ \rm km)^{-\frac{1}{3}}$, respectively.
These reconstructed H$_{\rm 0}$ values are consistent with the assumption on H$_{\rm 0} = 68$ km s$^{-1}$ Mpc$^{-1}$ within the errors.
This is probably because $A_{\tau} \tilde{F} G=$ 1 $({\rm pc}^2 \ \rm km)^{-\frac{1}{3}}$ was adopted when the mock FRBs were generated (Sect. \ref{simu_our_method}), and the two $A_{\tau} \tilde{F} G$ ranges cover this initial assumption reasonably well.
Therefore, in this ideal case, no significant systematics due to the prior $A_{\tau} \tilde{F} G$ assumption are expected.
However, this might not be the case when observed data are used to constrain H$_{\rm 0}$ because the physically reasonable range of $A_{\tau} \tilde{F} G$ is yet to be ascertained \citep{Cordes_2022}.
A further discussion of using observed data is described in Sect. \ref{AFG_for_obs_data}.

\subsection{Previous FRB method assuming \texorpdfstring{DM$_{\rm h}$}{DMh}}
\begin{figure}
        \centering
        \includegraphics[width=\columnwidth]{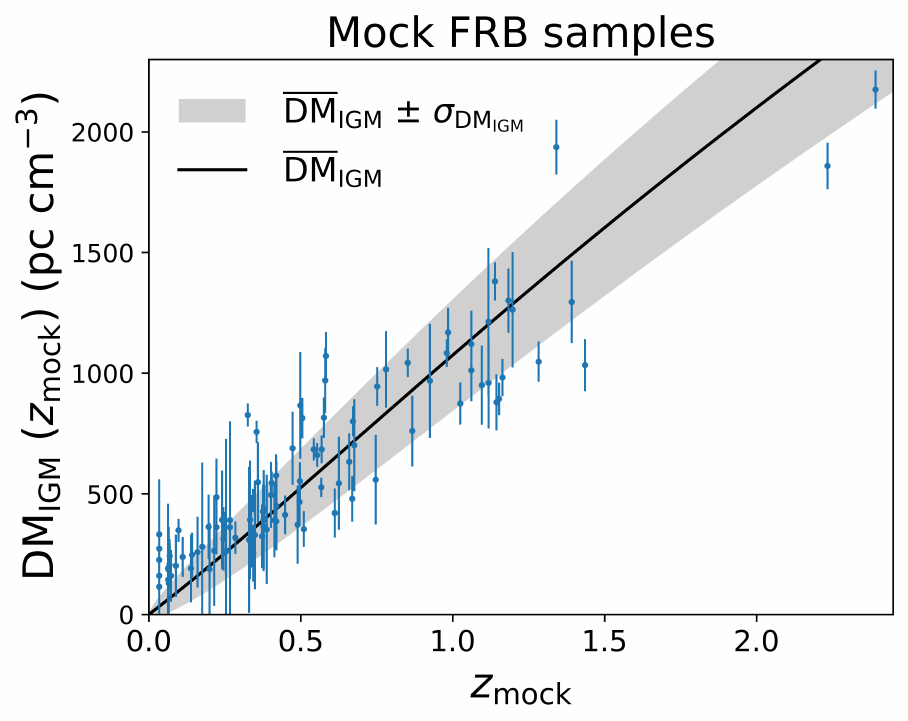}
        \includegraphics[width=\columnwidth]{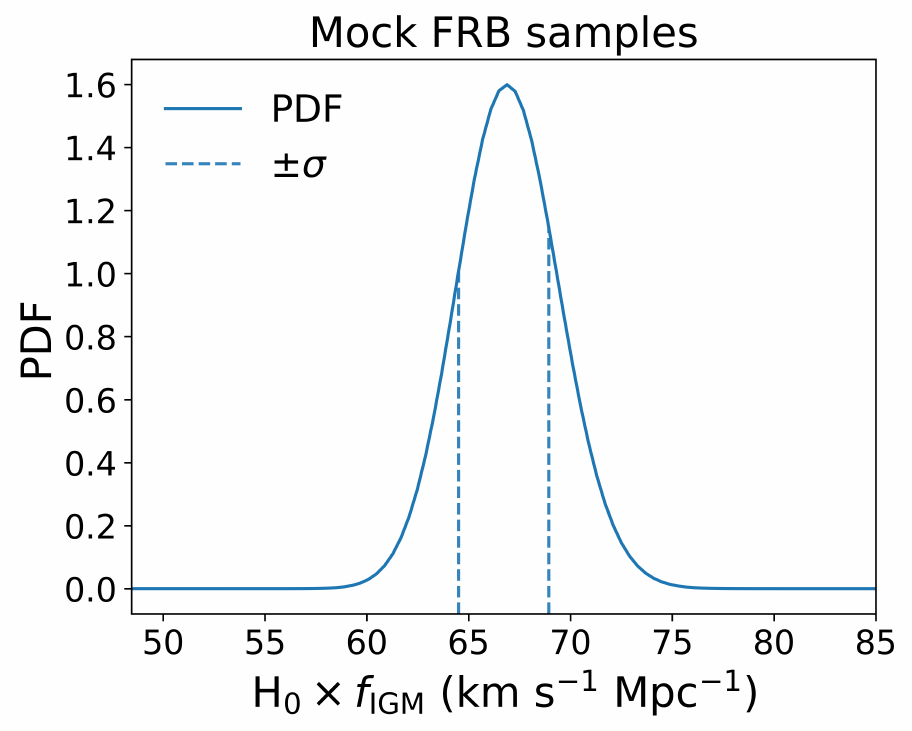}
        \includegraphics[width=\columnwidth]{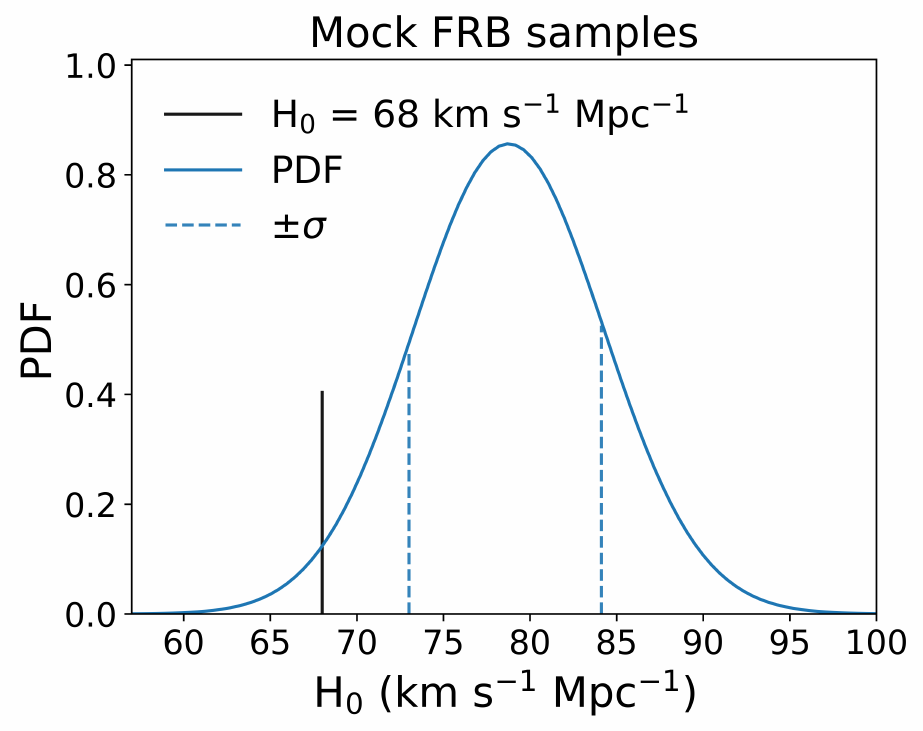}
        \caption{Simulations of the previous method with mock FRB samples. Top panel: DM$_{\rm IGM, mock}$ as a function of $z_{\rm mock}$.
        The blue dots with error bars show 100 mock FRB data, where DM$_{\rm IGM, mock}$ is calculated by the previous method, assuming DM$_{\rm h} = 50$ pc cm$^{-3}$.
        We fit Eq. \ref{gg} to the mock FRB data with a free parameter of H$_{\rm 0}$.
        The best-fit function is shown by the solid black line, where H$_{0}=78.7^{+5.4}_{-5.7}$ km s$^{-1}$ Mpc$^{-1}$. 
        The gray shaded region indicates the line-of-sight fluctuation of DM$_{\rm IGM}$ described by Eq. \ref{DM_IGM_e}.
        Middle panel: Same as the top panel of Fig. \ref{mock PDF}, but using the previous method, assuming DM$_{\rm h} = 50$ pc cm$^{-3}$.
        Bottom panel: Same as the bottom panel of Fig. \ref{mock PDF}, but using the previous method, assuming DM$_{\rm h} = 50$ pc cm$^{-3}$.
        }
    \label{mock PDF dmh}
    \end{figure}

    For a comparison, we tested a previous method without $\tau_{\rm mock, obs}$, assuming DM$_{\rm h}$ = 50 pc cm$^{-3}$ to calculate DM$_{\rm IGM}$. 
    In this previous method, DM$_{\rm IGM}$ was calculated solely based on Eq. \ref{DMeq} for the same 100 FRB mock samples as were used for our method above.
    The top panel of Fig. \ref{mock PDF dmh} illustrates the resulting DM$_{\rm IGM}$ as a function of $z_{\rm mock}$. 
    We fit Eq. \ref{gg} to the mock data by changing H$_{\rm 0}$ as a fitting parameter.
    The derived PDFs of H$_{0} \times f_{\rm IGM}$ and H$_{0}$ are shown in the middle and bottom panels of Fig. \ref{mock PDF dmh}, respectively.
    The result is H$_0$$\times f_{\rm IGM} = 66.9^{+2.4}_{-2.8}$ km s$^{-1}$ Mpc$^{-1}$.
    This value corresponds to H$_0 = 78.7^{+5.4}_{-5.7}$ km s$^{-1}$ Mpc$^{-1}$, where we assumed $f_{\rm IGM}=0.85$ taking the $\pm 0.05$ error of $f_{\rm IGM}$ into account. 
    This reconstructed H$_{\rm 0}$ significantly deviates from the assumed value of H$_{\rm 0} = 68.0$ km s$^{-1}$ Mpc$^{-1}$ in the simulation.

\subsection{Comparison between our method and the previous FRB method}
\begin{figure}
        \centering
        \includegraphics[width=\columnwidth]{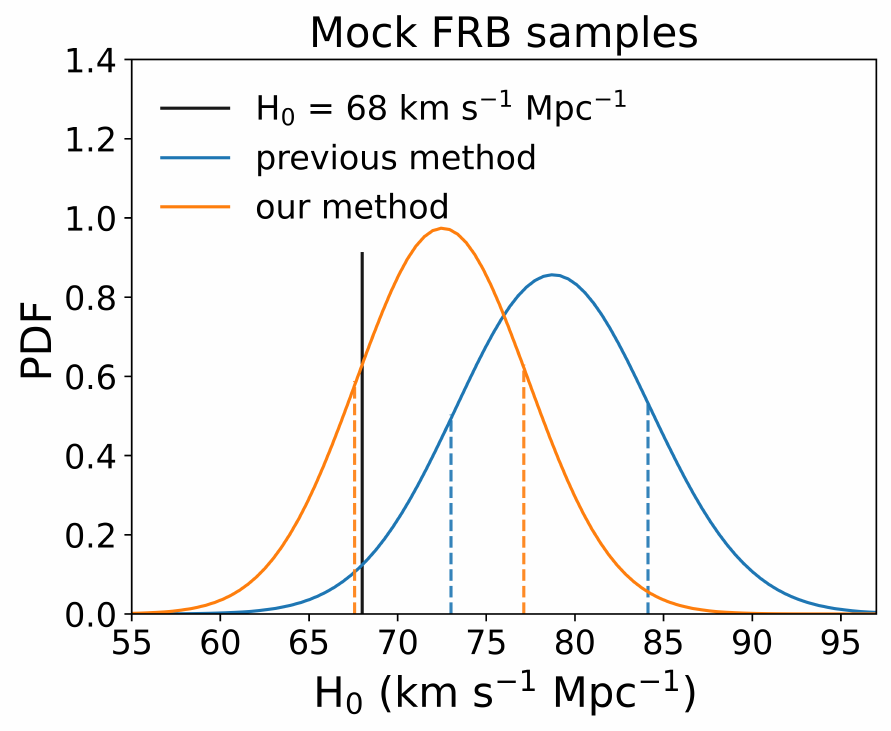}
        \caption{
        PDFs of H$_0$ derived by our method (orange) with positive and negative standard deviations (orange vertical dashed lines) and the previous method (blue) with positive and negative standard deviations (vertical dashed blue lines), using the same 100 mock FRB data.
        The vertical black line indicates H$_0 = 68.0$ km s$^{-1}$ Mpc$^{-1}$, which was assumed when we generated the 100 mock FRBs.
        }
    \label{cp mock PDF}
    \end{figure}

    Fig. \ref{cp mock PDF} summarizes the systematic difference between the previous method and our method.
    In Fig. \ref{cp mock PDF}, we note that the reconstructed H$_0$ value with our method is consistent with the assumed value of H$_{0}$ = 68 km s$^{-1}$Mpc$^{-1}$ within an uncertainty of 1 $\sigma$. The 1 $\sigma$ error is presented by the dashed vertical orange lines in Fig. \ref{cp mock PDF}, where the assumed value (solid vertical black line) is within this error. Therefore, the reconstructed H$_0$ with our method does not deviate in a statistically significant way from the assumed value. The apparent offset between our method and the assumed value in Fig. \ref{cp mock PDF} would be due to the statistical fluctuation owing to the random process in generating the mock FRB data.
    The systematic errors of H$_{\rm 0}$ are 15.7$\%$ and 6.6$\%$ for the previous method and our method, respectively.
    This corresponds to a reduction of the systematic error by 9.1\%.
    By adopting our method, the statistical error slightly decreases from 3.9\% to 2.9\%, which corresponds to a reduction of 1\%. 
    This small reduction is probably due to a systematic effect, where the intrinsic variation in DM$_{\rm h}$ contaminates the DM$_{\rm IGM}$ variation in the previous method, while our method reduces this contamination by measuring the individual DM$_{\rm h}$ from scattering.
    Notably, the reduction in the systematic error is closely aligned with the systematics of the Hubble tension, that is, an $\sim$10$\%$ systematic uncertainty. This highlights the potential of our method to address the Hubble tension using forthcoming FRB datasets.
\section{Selection criteria for the observed FRB samples}
\label{samples}
    In the previous section, we applied our method to the simulated mock FRB samples to demonstrate how our method works compared with the previous method.
    In this section, we apply our method as described in Sect. \ref{S:data} to 30 localized FRBs, which are shown in Table \ref{cat}. 
    In our method, we require localized FRBs with measurements of the spectroscopic redshift.
    Although more than 700 FRBs have been detected so far\footnote{\url{https://blinkverse.alkaidos.cn/}}, only $\sim$40 of them are localized to host galaxies \citep[e.g.,][]{Bhandari_2022,Law2023}. 
    Additionally, our method requires information about the scattering time, which meant that we were left with 37 data sets to work with. 
    According to \citet{Cordes_2022}, the reasonable range of the $A_{\tau} \tilde{F} G$ parameter is smaller than 10 $({\rm pc}^2 \ {\rm km})^{-\frac{1}{3}}$.
    The PDF of the $A_{\tau} \tilde{F} G$ parameter can be computed by integrating Eq. \ref{3D} (the top panels of Fig. \ref{fdmh}) over DM$_{\rm h}$.
    We found that the PDFs of the $A_{\tau} \tilde{F} G$ parameter peak beyond 10 $({\rm pc}^2 \ {\rm km})^{-\frac{1}{3}}$ for four FRB samples, that is, FRBs 20191228A, 20210405I, 20210410D, and 20220912A.
    Their PDFs of the $A_{\tau} \tilde{F} G$ parameter are mostly distributed at $A_{\tau} \tilde{F} G$ $>$ 10 $({\rm pc}^2 \ {\rm km})^{-\frac{1}{3}}$.
    The range of the peaks is $A_{\tau} \tilde{F} G = [10,1000]\ ({\rm pc}^2 \ {\rm km})^{-\frac{1}{3}}$ for these four samples. 
    Therefore, we excluded these four FRBs from our samples as outliers in the $A_{\tau} \tilde{F} G$ parameter.

\begin{figure}
        \centering
        \includegraphics[width=\columnwidth]{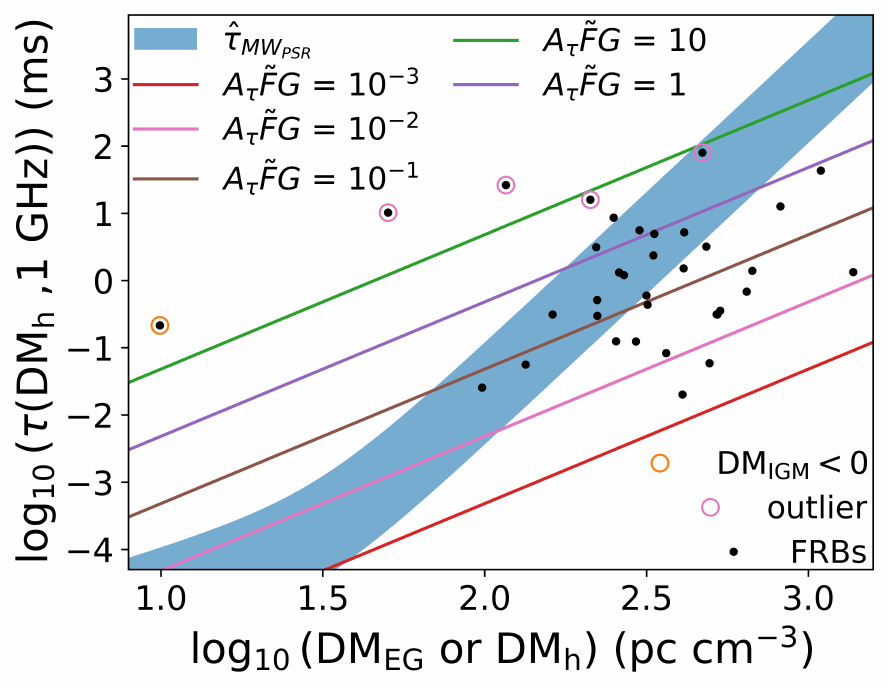}
        \caption {Scattering time vs. ${\rm DM_{EG}}$ or DM$_{h}$. 
        The $x$-axis is the ${\rm DM_{EG}}$ from Eq. (\ref{pipi}) in the log scale for the FRB samples (black dots), and it shows DM$_{\rm h}$ for the empirical relation based on Galactic pulsars (shaded blue region) and the $A_{\tau} \tilde{F} G$ models (solid colored lines). 
        The $y$-axis is the scattering time at the rest-frame frequency of 1 GHz in the log scale. 
        The black points are the 30 FRBs in our work. 
        The pink circles are the $A_{\tau} \tilde{F} G$ outliers in our FRB samples.
        The orange circle is one of our FRB samples, which is DM$_{\rm IGM}<0$ pc cm$^{-3}$.
        The colored lines are the $A_{\tau} \tilde{F} G$ models from Eq. (\ref{yaya}) with ${\rm DM_h}=[20,1600]$ pc cm$^{-3}$. 
        The blue area is the empirical relation derived from Galactic pulsars described in Eq. (\ref{jjj}), which includes the geometrical scaling factor for the FRB samples. 
        }
        \label{DM_EG-tau}
    \end{figure}

    We also excluded three FRB samples that had negative DM$_{\rm IGM}$ as follows.
    The extragalactic component of the dispersion measure (${\rm DM_{EG}}$) is described as
    \begin{equation}
        DM_{\rm EG} \equiv {\rm DM}_{\rm IGM}+ \frac{\rm DM_h}{1+z_{\rm spec}} = {\rm DM_{\rm obs} - DM_{MW_{disk}} - DM_{MW_{halo}}},
    \label{pipi}
    \end{equation}
    where ${\rm DM_{MW_{halo}}}=52.5$ pc cm$^{-3}$, and ${\rm DM_{MW_{disk}}}$ was adopted from the NE2001 model. 
    Here, certain values were used for ${\rm DM_{MW_{halo}}}$ and ${\rm DM_{MW_{disk}}}$ for the sample selection alone. 
    The PDFs of ${\rm DM_{MW_{halo}}}$ and ${\rm DM_{MW_{disk}}}$ were taken into account in our method (Sect. \ref{S:data}) to constrain H$_{\rm 0}$.
    We found that DM$_{\rm EG}$ of FRB 20200120E and 20220319D are negative, indicating a negative DM$_{\rm IGM}$.
    The ${\rm DM_{EG}}$ of FRB 20181030A is 9.93 pc cm$^{-3}$. 
    This value is lower than the lower bound of the prior assumption on DM$_{\rm h}$ ($=20$ pc cm$^{-3}$) adopted in this work, following \citet{Cordes_2022}.
    These three FRBs correspond to negative DM$_{\rm IGM}$.
    Therefore, we also excluded these three FRBs from our analysis.
    We excluded 7 of the 37 data sets, and the remaining 30 FRB samples were used in this work. 
    Table \ref{cat2} lists the 7 FRBs we excluded from our samples. 
    In Fig. \ref{DM_EG-tau}, we compare the scattering time at the rest-frame 1 GHz and DM$_{\rm EG}$ for the 30 FRB samples, along with 5 samples out of 7 excluded samples.
    We only show 5 excluded samples because 2 of the 7 have a negative DM$_{\rm EG}$.
    These 5 excluded samples are close to or above the model line of $A_{\tau} \tilde{F} G = 10$ $({\rm pc}^2 \ {\rm km})^{-\frac{1}{3}}$ (solid green line in Fig. \ref{DM_EG-tau}).
    The empirical relation between the scattering at the rest-frame 1 GHz and DM$_{\rm h}$ expected from Galactic pulsar observations is described as
    \begin{equation}
    \begin{aligned}
        &\log_{10}(\tau_{\rm MW_{psr}}) =\\& \log_{10}(1.9 \times 10^{-7}\ {\rm ms}\ {\rm DM_h}^{1.5} \times (1 + 3.55\times 10^{5}\ {\rm DM_h}^{3}))\\& +\log_{10}(3)\pm 0.76,
    \label{jjj}
    \end{aligned}
    \end{equation}
    \citep{Cordes_2022}, which is shown as the shaded blue region in Fig. \ref{DM_EG-tau}.
    The term of $\log_{10}(3)$ represents a scaling factor for taking the spherical wavefront difference between pulsars in the Milky Way and FRBs in host galaxies in the extragalactic Universe into account. 
    \begin{table*}
        \centering
        \renewcommand{\arraystretch}{1.5}
        \caption{FRB samples.}
        \begin{tabular}{lccccccccc}
            \hline
             FRB & \multicolumn{1}{c}{DM$_{\rm obs}$} & ${\rm DM_{MW_{disk,obs}}}$ & $v_{\rm obs}$ & \multicolumn{1}{c}{$z_{\rm spec}$} & $\tau_{\rm obs}$ & $\sigma_{\tau_{\rm obs}}$& \multicolumn{3}{c}{References} \\ 
             & \multicolumn{1}{c}{(pc cm$^{-3}$)} & (pc cm$^{-3}$) & (GHz) &  & (ms) & (ms) & DM$_{\rm obs}$ & $z_{\rm spec}$ & $\tau_{\rm obs}$, $v_{\rm obs}$\\
            \hline
            20121102A & $557.4^{+2.0}_{-2.0}$ & 188  & 0.5   & $0.19273\pm 0.0008$ & $<9.6$  &         & 1 & 20  & 23 \\
            20180301A & $536^{+8}_{-13}$ & 152  & 1.35  &  0.3304 & 0.71    & 0.03    & 2 & 2  & 24 \\
            20180916B & $349.2^{+0.3}_{-0.3}$ & 198  & 0.35  &  0.0337 & $<1.7$  &         & 3 & 2  & 25 \\
            20180924A & $361.42^{+0.06}_{-0.06}$ & 40.5 & 1.2   & $0.3214\pm 0.0002$ & 0.68    & 0.03    &  4 & 5  & 9 \\
            20181112A & $589.27^{+0.03}_{-0.03}$ & 41.7 & 1.3   & $0.47550\pm 0.00015$ & 0.0207  & 0.00085 & 5 & 5  & 26 \\
            20190102C & $363.6^{+0.3}_{-0.3}$ & 57   & 1.32  & 0.2913 & 0.041   & 0.0025  & 6 & 6  & 9 \\
            20190520B & $1204.7^{+0.4}_{-0.4}$ & 60.2 & 1.41  & $0.241\pm 0.001$ & 10.9    & 1.5     & 7 & 7  & 27 \\
            20190523A & $760.8^{+0.6}_{-0.6}$ & 37.2 & 1     & $0.660\pm 0.002$ & 1.4     & 0.2     & 8 & 8  & 8 \\
            20190608B & $340.05^{+0.06}_{-0.03}$ & 37.3 & 1.27  & 0.11778 & 3.3     & 0.2     & 9 & 9  & 9 \\
            20190611B & $321.4^{+0.2}_{-0.2}$ & 57.8 & 1.3   & 0.378 & 0.18    & 0.02    & 6 & 9  & 9 \\
            20190614D & $959.2^{+5}_{-5}$ & 87.8 & 1.4   & $0.60\pm 0.17$ & $<3.3$  &         & 10 & 10  &  10\\
            20190711A & $593.1^{+0.4}_{-0.4}$ & 56.5 & 1.3   & 0.522 & $<1.12$ &         & 6 & 9  & 28 \\
            20190714A & $504^{+2}_{-2}$ & 38.5 & 1.27  & 0.2365 & $<2$    &         & 11 & 21  & 11 \\
            20191001A & $506.92^{+0.04}_{-0.04}$ & 44.2 & 0.824 & $0.2340\pm0.0001$ & 3.3     & 0.02    & 12 & 12  & 12 \\
            20200430A & $380.1^{+0.4}_{-0.4}$ & 27.2 & 0.865 & 0.1610 & $<15$   &         & 13 & 21  & 13 \\
            20201124A & $413.52^{+0.05}_{-0.05}$ & 140  & 0.865 & $0.098\pm 0.002$ & $<9$    &         & 14 & 14 & 29 \\
            20210117A & $729.1^{+0.36}_{-0.23}$ & 34.4 & 1.2   & $0.214\pm 0.001$ & 0.33    & 0.02    & 15 & 21  & 30 \\
            20210320C & $384.8^{+0.3}_{-0.3}$ & 39.3 & 0.842 & 0.280 & 0.247   & 0.006   & 16 & 16 & 30 \\
            20210603A & $500.147^{+0.004}_{-0.004}$ & 39.5 & 0.6   & $0.1772\pm 0.0001$ & 0.155   & 0.003   & 17 & 17 & 17 \\
            20220207C & $262.38^{+0.01}_{-0.01}$ & 76.1 & 1.4   & 0.043040 & 0.0146  & 0.0026  & 18 & 18 & 18\\
            20220307B & $499.27^{+0.06}_{-0.06}$ & 128  & 1.4   & 0.248123 & 0.1139  & 0.0059  & 18 & 18 & 18\\
            20220310F & $462.24^{+0.005}_{-0.005}$ & 46.3 & 1.4   & 0.477958 & 0.02164 & 0.00022 & 18 & 18 & 18\\
            20220418A & $623.25^{+0.01}_{-0.01}$ & 36.7 & 1.4   & 0.622000 & 0.0928  & 0.003   & 18 & 18 & 18\\
            20220506D & $396.97^{+0.02}_{-0.02}$ & 84.6 & 1.4   & 0.30039 & 0.3427  & 0.0074  & 18 & 18 & 18\\
            20220509G & $269.53^{+0.02}_{-0.02}$ & 55.6 & 1.41  & 0.089400 & 0.08    & 0.02    & 18 & 22 & 22\\
            20220610A & $1458.15^{+0.25}_{-0.55}$ & 31   & 1.15  & $1.016\pm 0.002$ & 0.511   & 0.012   & 19 & 19 & 19\\
            20220825A & $651.24^{+0.06}_{-0.06}$ & 78.5 & 1.4   & 0.241397 & 0.0817  & 0.0033  & 18 & 18 & 18\\
            20220914A & $631.28^{+0.04}_{-0.04}$ & 54.7 & 1.4   & 0.113900 & $<0.08$ &         & 18 & 22 & 22\\
            20220920A & $314.99^{+0.01}_{-0.01}$ & 39.9 & 1.4   & 0.158239 & 0.078   & 0.15    & 18 & 18 & 18\\
            20221012A & $441.08^{+0.7}_{-0.7}$ & 54.3 & 1.4   & 0.284669 & $<1.29$ &         & 18 & 18 & 18\\
            \hline
        \end{tabular}
        \tablebib{(1)~\citet{Spitler+2014}; (2) \citet{Bhandari_2022}; (3) \citet{CHIME/FRBCollaboration+2019a}; (4) \citet{Bannister2019}; (5) \citet{Prochaska+2019}; (6) \citet{Macquart_2020}; (7) \citet{2022SciA....8I6375N}; (8) \citet{2019Natur.572..352R}; (9) \citet{10.1093/mnras/staa2138}; (10) \citet{Law_2020}; (11) \citet{2019ATel12940....1B}; (12) \citet{Bhandari_2020}; (13) \citet{2020ATel13694....1K}; (14) \citet{10.1093/mnras/stac465}; (15) \citet{Bhandari2023}; (16) \citet{Shannon2023}; (17) \citet{cassanelli2023fast}; (18) \citet{Law2023}; (19) \citet{ryder2022probing}; (20) \citet{Tendulkar+2017}; (21) \citet{simha2023searching}; (22) \citet{Connor_2023}; (23) \citet{Josephy_2019}; (24) \citet{Price_2019}; (25) \citet{2020ApJ...896L..41C}; (26) \citet{Cho_2020}; (27) \citet{Lee_2023}; (28) \citet{10.1093/mnras/staa1916}; (29) \citet{2021ATel14502....1K}; (30) \citet{sammons2023twoscreen}.}
        \label{cat}
    \end{table*}
    
    \begin{table*}
        \centering
        \renewcommand{\arraystretch}{1.5}
        \caption{DM$_{\rm h}$ and $z_{\rm model}$ computed for the FRB samples.}
        \begin{tabular}{lcc|lcc}
            \hline
             FRB & ${\rm DM_h}\ ({\rm H}_{\rm 0}=73)$ & $z_{\rm model}$ & FRB & ${\rm DM_h}\ ({\rm H}_{\rm 0}=73)$ & $z_{\rm model}$\\
             & pc cm$^{-3}$ &  &  & pc cm$^{-3}$ &  \\
            \hline
            20121102A & $58^{+25}_{-60}$  & $0.301\pm 0.13$ & 20201124A & $114^{+48}_{-54}$ & $0.15\pm 0.1$ \\
            20180301A & $141^{+72}_{-56}$ & $0.251\pm 0.1$ & 20210117A & $118^{+112}_{-64}$ & $0.602\pm 0.15$ \\
            20180916B & $33^{+10}_{-23}$ & $0.1\pm 0.05$   & 20210320C & $48^{+44}_{-19}$ & $0.326\pm 0.1$ \\
            20180924A & $97.5^{+59}_{-43}$ & $0.226\pm 0.1$ & 20210603A & $37.1^{+46}_{-13}$ & $0.451\pm 0.13$ \\
            20181112A & $85.6^{+80}_{-45}$ & $0.501\pm 0.15$ & 20220207C & $64.1^{+31}_{-29}$ & $0.125\pm 0.075$ \\
            20190102C & $75.2^{+52}_{-37}$ & $0.251\pm 0.1$ & 20220307B & $91.4^{+65}_{-47}$ & $0.301\pm 0.13$ \\
            20190520B & $827^{+199}_{-161}$ & $0.576\pm 0.18$ & 20220310F & $89.1^{+70}_{-47}$ & $0.351\pm 0.13$ \\
            20190523A & $178^{+133}_{-57}$ & $0.576\pm 0.18$ & 20220418A & $104^{+91}_{-57}$ & $0.526\pm 0.15$ \\
            20190608B & $199^{+33}_{-32}$ & $0.1\pm 0.085$   & 20220506D & $56.9^{+56}_{-49}$ & $0.226\pm 0.1$ \\
            20190611B & $74.4^{+46}_{-36}$ & $0.226\pm 0.1$ & 20220509G & $71.7^{+37}_{-33}$ & $0.15\pm 0.075$ \\
            20190614D & $297^{+177}_{-168}$ & $0.652\pm 0.2$ & 20220610A & $283^{+240}_{-101}$ & $1.28\pm 0.25$ \\
            20190711A & $133^{+103}_{-72}$ & $0.401\pm 0.15$ & 20220825A & $104^{+92}_{-57}$ & $0.501\pm 0.15$ \\
            20190714A & $148^{+90}_{-79}$ & $0.326\pm 0.15$ & 20220914A & $56.3^{+74}_{-25}$ & $0.501\pm 0.15$ \\
            20191001A & $143^{+89}_{-44}$ & $0.326\pm 0.13$ & 20220920A & $75.9^{+47}_{-37}$ & $0.226\pm 0.075$ \\
            20200430A & $154^{+64}_{-75}$ & $0.201\pm 0.13$ & 20221012A & $134^{+75}_{-69}$ & $0.251\pm 0.13$ \\
            
            \hline
        \end{tabular}
        \label{cat10}
    \end{table*}
    \begin{table*}
        \centering
        \caption{FRB samples that we excluded from our analysis either because of (i) outliers in the $A_{\tau} \tilde{F} G$ parameter space or because (ii) the DM$_{\rm EG}$ is negative.}
        \begin{tabular}{lccccccccc}
            \hline
             FRB & DM$_{\rm obs}$ & ${\rm DM_{MW_{disk,obs}}}$ & $v_{\rm obs}$ & $z_{\rm spec}$ & $\tau_{\rm obs}$ & $\sigma_{\tau_{\rm obs}}$ & $\tau (1\ {\rm GHz})$ & ${\rm DM_{EG}}$ & \multicolumn{1}{c}{References} \\ \\
             & pc cm$^{-3}$ & pc cm$^{-3}$ & GHz &  & ms & ms & ms & pc cm$^{-3}$ & DM, $z_{\rm spec}$, $\tau_{\rm obs}$, $v_{\rm obs}$\\
            \hline
            20191228A & 298  & 33   & 1.27  & 0.2432  & 6.1        & 0.6  & 15.9    & 212   & 1 \\
            20210405I & 565  & 396  & 1.28  & 0.066   & 9.7        & 0.2  & 26.4    & 116   & 2 \\
            20210410D & 579  & 56.2 & 1.28  & 0.1415  & 29.4       & 2    & 79.9    & 470   & 3 \\
            20220912A & 228  & 125  & 1.41  & 0.0771  & 2.63       & 0.35 & 10.3    & 50.3  & 4 \\
            20181030A & 104  & 41.1 & 0.704 & 0.0039  & $<0.88$    &      & 0.22    & 9.93  & 5 \\
            20200120E & 87.8 & 41   & 1.4   & 0.0008  & $<0.00003$ &      & 0.00011 & -5.63 & 6 \\
            20220319D & 111  & 140  & 1.4   & 0.01123 & $<0.1$     &      & 0.38    & -81.3 & 7 \\
            \hline
        \end{tabular}
        \tablebib{(1)~\citet{Bhandari_2022}; (2) \citet{driessen2023frb}; (3) \citet{Caleb_2023}; (4) \citet{Ravi_2023}; (5) \citet{Amiri_2021}; (6) \citet{nimmo2021burst}; (7) \citet{ravi2023deep}.
        }
        \label{cat2}
    \end{table*}

\section{Results for the observed FRB samples}\label{sec:result}
\begin{figure}
        \centering
        \includegraphics[width=0.93\columnwidth]{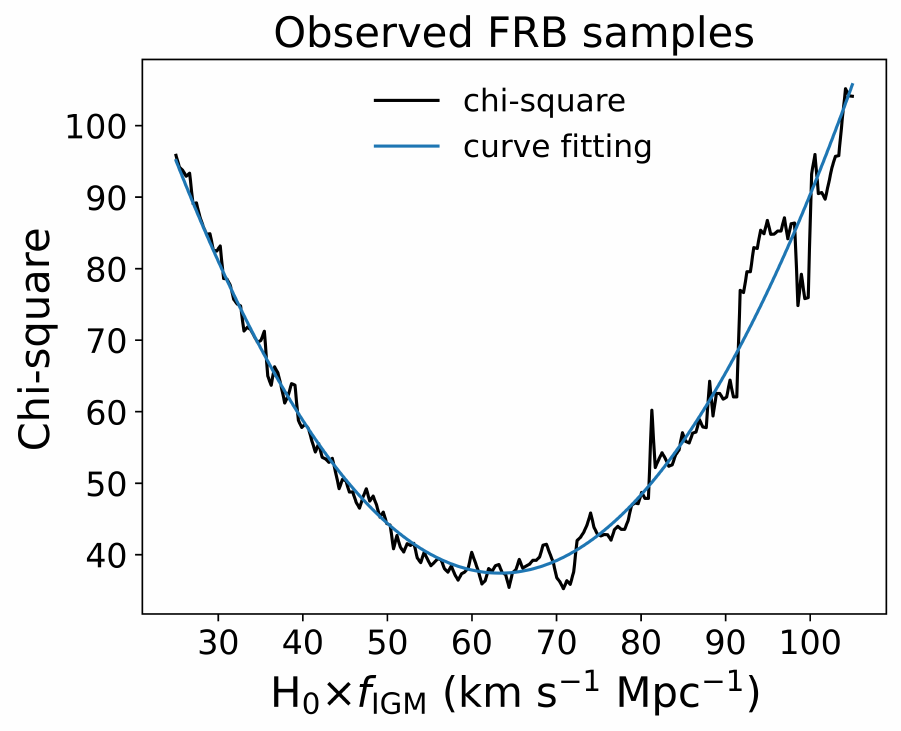}
        \includegraphics[width=0.93\columnwidth]{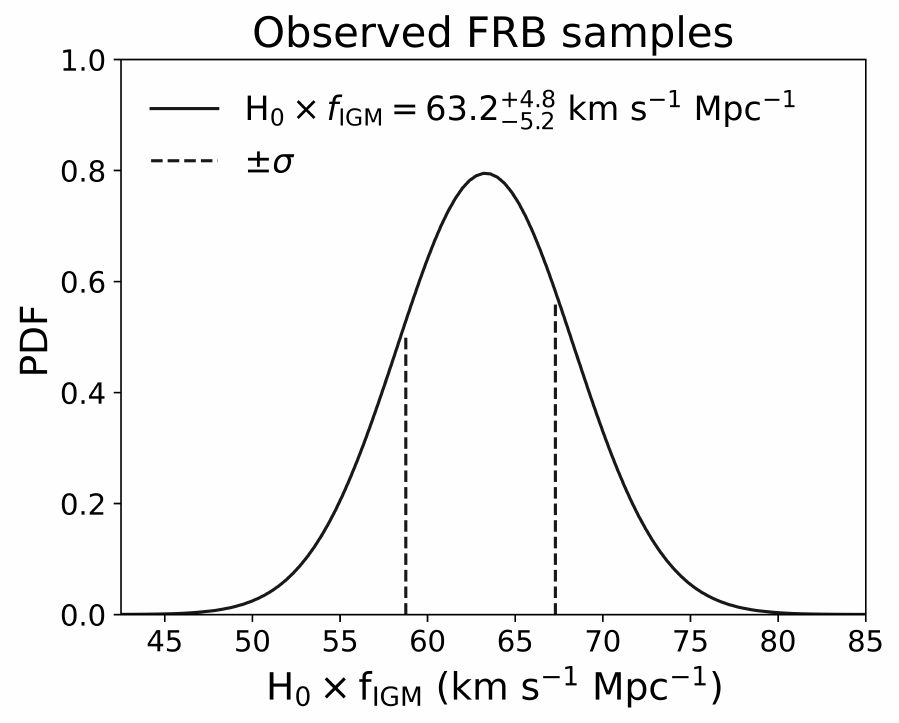}
        \includegraphics[width=0.93\columnwidth]{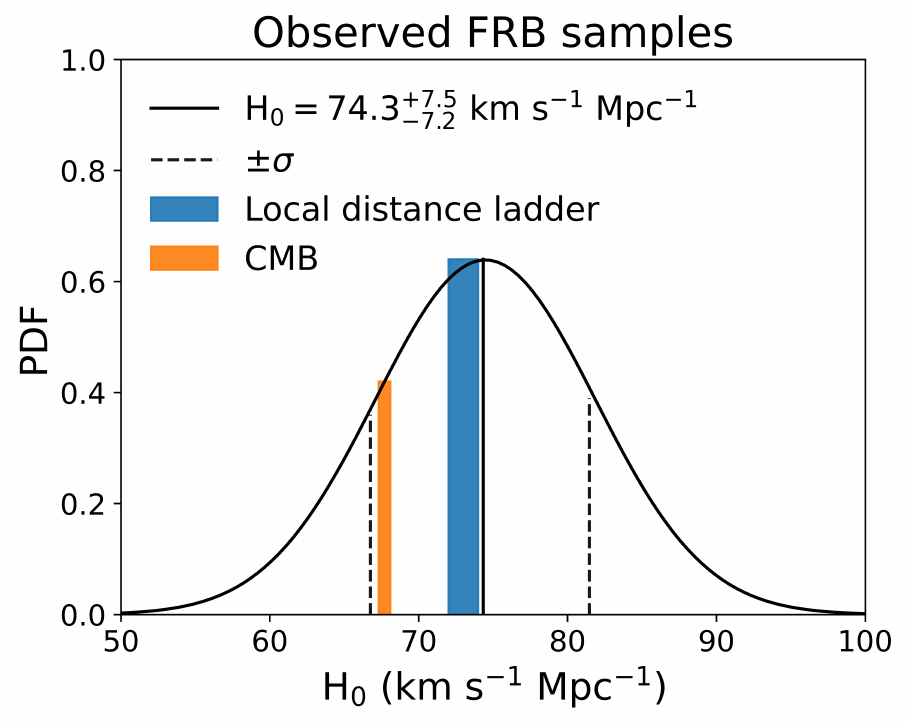}
        
        \caption {The results of our method observed FRB samples. Top panel: $\chi^{2}$ described by Eq. \ref{chi2} as a function of $f_{\rm IGM}\times {\rm H}_{\rm 0}$ using the 30 FRB samples and our method. 
        The black line indicates the $\chi^{2}$ values derived by changing $f_{\rm IGM}\times {\rm H}_{\rm 0}$. 
        The blue line is the best-fit polynomial function to $\chi^{2}$. Middle panel: PDF of $f_{\rm IGM}\times {\rm H}_{\rm 0}$ from the  $\chi^{2}$ (top panel). The vertical dashed black lines correspond to the 84.2 and 15.8 percentiles ($\pm \sigma$) of the PDF. Bottom panel: PDF of H$_0$ by changing the scale from H$_{\rm 0}\times f_{\rm IGM}$ (middle panel) to H$_0$ for a given $f_{\rm IGM}=0.85\pm0.05$. The solid vertical line indicates the peak of the PDF, and the dashed vertical lines indicate the uncertainty range of H$_{\rm 0}$ after taking the 0.05 error of $f_{\rm IGM}$ into account.
        The orange area indicates the CMB measurement of ${\rm H}_{\rm 0} $ \citep[e.g.,][]{refId0}. 
        The blue area shows the measurement by local distance ladders \citep[e.g.,][]{Riess_2022}.}
        \label{88}
    \end{figure}

\begin{figure}
        \centering
        \includegraphics[width=\columnwidth]{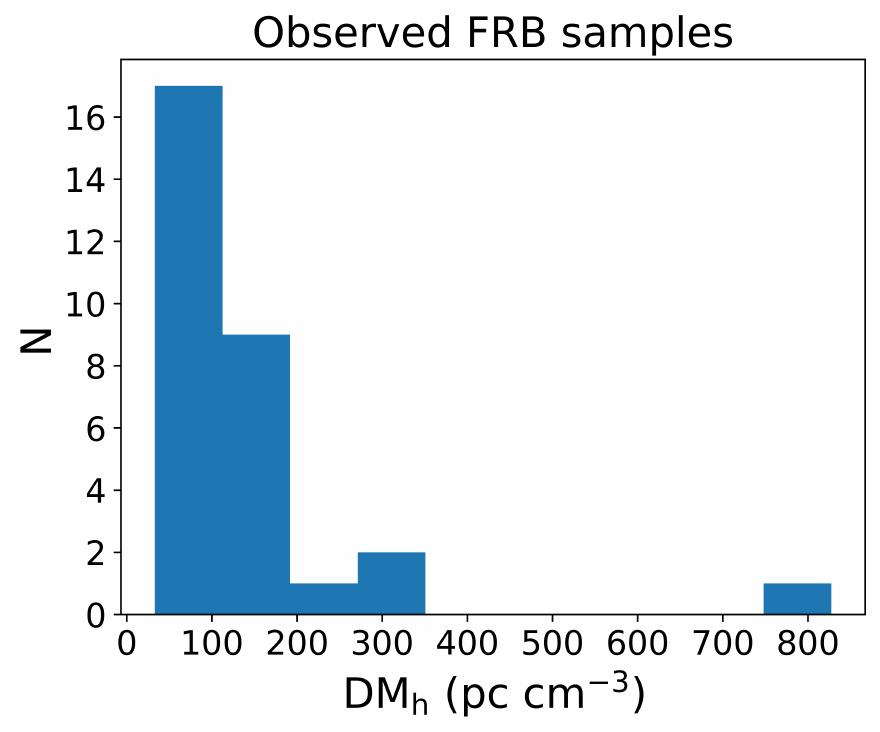}
        
        \caption {Histogram of $\rm DM_h$ derived by using scattering (see also Table \ref{cat10}).}
        \label{hisDMh}
    \end{figure}

    As described in Sect. \ref{opt_H0}, we treated $f_{\rm IGM} \times {\rm H}_{\rm 0}$ as a single parameter to fit $z_{\rm model}$ to $z_{\rm spec}$.
    The top panel of Fig. \ref{88} shows the $\chi^{2}$ of the fitting as a function of $f_{\rm IGM} \times {\rm H}_{\rm 0}$,
    \begin{equation}
        \chi^{2} = \sum_{i} \frac{(z_{\rm model, i}-z_{\rm spec, i})^{2}}{\sigma_{i}^{2}},
    \label{chi2}
    \end{equation}
    where $\sigma_{i}$ is the error of $z_{\rm model}$ of the $i$th FRB sample.
    We calculated the PDF of $f_{\rm IGM} \times {\rm H}_{\rm 0}$ assuming PDF $\propto \exp({-\chi^{2}/2}$).
    The result is shown in the middle panel of Fig. \ref{88} with prior assumptions on ${\rm DM_h} = [20, 1600]$ pc cm$^{-3}$, and $A_{\tau} \tilde{F} G = [0.001,10]$ $({\rm pc}^2 \ {\rm km})^{-\frac{1}{3}}$.
    The best-fit result is $f_{\rm IGM}\times {\rm H}_{\rm 0}=63.2^{+4.8}_{-5.2}$ km s$^{-1}$ Mpc$^{-1}$.
    Given $f_{\rm IGM}=0.85\ \pm 0.05$ \citep{Cordes_2022}, the best-fit result corresponds to ${\rm H}_{\rm 0}=74.3^{+7.2}_{-7.5}$ km s$^{-1}$ Mpc$^{-1}$, where the uncertainty of $f_{\rm IGM}$ was taken into account (bottom panel of Fig. \ref{88}). 
    The median value of ${\rm DM_h}$ in our analysis is DM$_{\rm h}$ $=103^{+68}_{-48}$ pc cm$^{-3}$ (see also Table \ref{cat10} and Fig. \ref{hisDMh}).
    These errors were determined by calculating the median of the ${\rm DM_h}$ errors.

\section{Discussion}\label{discussion}
\subsection{Comparison with the CMB and local distance ladders}
Our measurement of the Hubble constant is ${\rm H}_{\rm 0}=74.3 ^{+7.2} _{-7.5}$  km s$^{-1}$ Mpc$^{-1}$ using scattering. 
The central value of this measurement prefers the measurement from the local distance ladder \citep[H$_{0}= 73.0 \pm 1.0$ km s$^{-1}$ Mpc$^{-1}$;][]{Riess_2022} than the CMB \citep[H$_{0}= 67.7 \pm 0.4$ km s$^{-1}$ Mpc$^{-1}$;][]{refId0}. 
However, our measurement is still consistent with these two methods within the 1 $\sigma$ error.
To address the Hubble tension with FRBs, both statistical and systematic errors have to be reduced in the future, as discussed in the following sections.

\subsection{Comparison with the other FRB methods}
\label{Compare paper}
    Recently, some papers reported ${\rm H}_{\rm 0}$ values that were constrained by FRBs, and the methods differed from ours. 
    \citet{Hagstotz_2022} assumed a normal distribution of DM$_{\rm h}$ with $\mu=$ 100 pc cm$^{-3}$ and $\sigma =$ 50 pc cm$^{-3}$.
    They derived ${\rm H}_{\rm 0}=62.3\pm 9.1$ km s$^{-1}$ Mpc$^{-1}$ using the DM$_{\rm IGM}$–$z$ relation with nine localized FRB samples.
    They assumed $f_{\rm IGM}=0.84$, where the difference from our assumption, $f_{\rm IGM}=0.85$, is negligibly small compared to the uncertainty of H$_{\rm 0}$. 
    \citet{James_2022} combined FRB population models and the DM$_{\rm EG}$-$z$ relation, which were parameterized by seven quantities for fitting, including $\mu$ and $\sigma$ of the lognormal distribution of DM$_{\rm h}$ and H$_{0}$.
    They fit the model to the 76 FRB data obtained from The Commensal Real-time The Australian Square Kilometre Array Pathfinder Fast
Transients Coherent (CRACO) survey (16 localized and 60 unlocalized), where the best-fit model indicates ${\rm H}_{\rm 0}=73^{+12}_{-8}$ km s$^{-1}$ Mpc$^{-1}$, $\mu=186^{+59}_{-48}$ pc cm$^{-3}$, and $\sigma=3.5$ pc cm$^{-3}$. They assumed $f_{\rm IGM}=0.844$. 
    \citet{2022arXiv221213433Z} used a Bayesian framework to estimate H$_0$ = 
    80.4 $^{+24.1}_{-19.4}$ km s$^{-1}$ Mpc$^{-1}$ with 12 unlocalized FRB samples with 
    a prior assumption on a lognormal distribution of DM$_{\rm h}$ with $\mu$ = 68 pc 
    cm$^{-3}$ and $\sigma$ = 0.88 pc cm$^{-3}$.
    The samples were collected from FRBs detected with the Australian Square Kilometre Array Pathfinder (ASKAP).
    These previous works assumed a certain shape of the DM$_{\rm h}$ distribution, that is, a lognormal distribution or a normal distribution.
    Our method is free of this assumption. The scattering is used to derive the individual DM$_{\rm h}$ and DM$_{\rm IGM}$ in the samples. 
    
    Our result is ${\rm H}_{\rm 0}=74.3^{+7.2}_{-7.5}$ km s$^{-1}$ Mpc$^{-1}$ (Sect. \ref{sec:result}), which is consistent with H$_{\rm 0}$ in these previous works within the statistical errors.
    In the previous method, however, there might be significant unknown systematics in the assumption on DM$_{\rm h}$ as we demonstrated in Sect. \ref{simulation}.
    Depending on the different assumptions on DM$_{\rm h}$, the central values of derived H$_{0}$ systematically differ in the previous works mentioned above.
    This possible systematics is still smaller than the large statistical error based on the current FRB samples. 
    However, the possible systematics would be problematic in the future when the statistical uncertainty is significantly reduced by large FRB samples.
    In contrast to previous studies, our method can minimize these systematics using the scattering time to derive the individual DM$_{\rm h}$ rather than assuming a certain DM$_{\rm h}$ distribution.

\subsection{Prior assumption on the \texorpdfstring{$A_{\tau} \tilde{F} G$}{AFG} parameter range}
\label{AFG_for_obs_data}
    Our simulation described in Sect. \ref{priorAFG} suggests that the prior assumption on the $A_{\tau} \tilde{F} G$ range does not significantly impact the best-fit result of H$_{0}$, as far as the prior covers the true $A_{\tau} \tilde{F} G$ value.
    However, this might not be the case when observed data are used because the true $A_{\tau} \tilde{F} G$ distribution is unknown \citep{Cordes_2022}.
    
\begin{figure}
        \centering
        \includegraphics[width=\columnwidth]{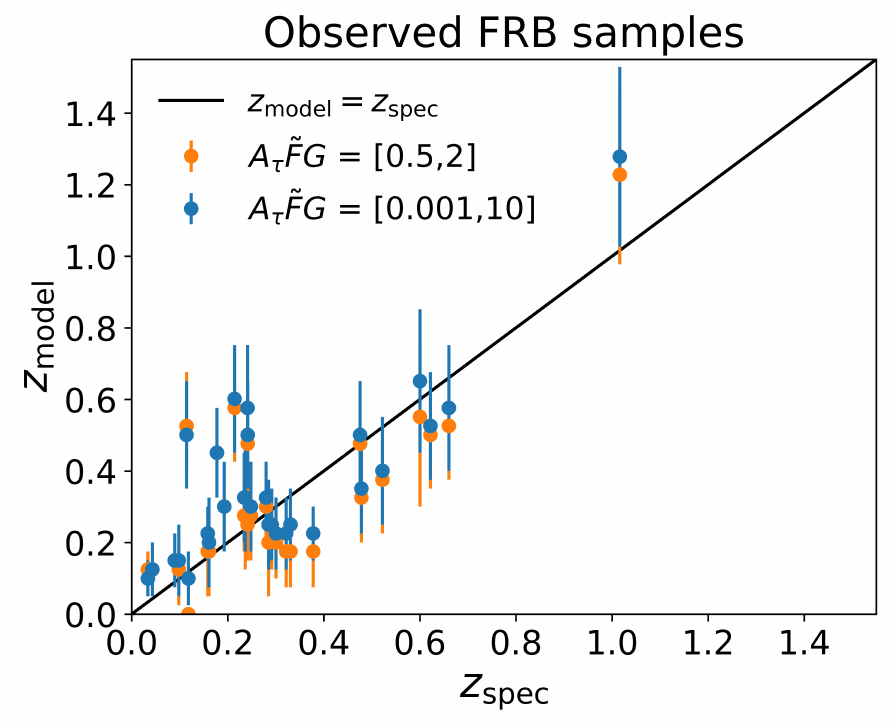}
        
        \caption {$z_{\rm model}$ vs. $z_{\rm obs}$, comparing two prior assumptions on the $A_{\tau} \tilde{F} G$ ranges using the 30 FRB samples and our method.
        The blue points with error bars assume $A_{\tau} \tilde{F} G$ $=[0.001,10]$ $({\rm pc}^2 \ {\rm km})^{-\frac{1}{3}}$. 
        The orange points with error bars assume $A_{\tau} \tilde{F} G$ $=[0.5,2]$ $({\rm pc}^2 \ {\rm km})^{-\frac{1}{3}}$.
        To demonstrate how the result depends on the prior assumption on the $A_{\tau} \tilde{F} G$ range, the identical value of $f_{IGM}\times {\rm H}_{\rm 0}=63.2$ km s$^{-1}$ Mpc$^{-1}$ is adopted for both cases. 
        }
        \label{878}
    \end{figure}

\begin{figure}
        \centering
        \includegraphics[width=\columnwidth]{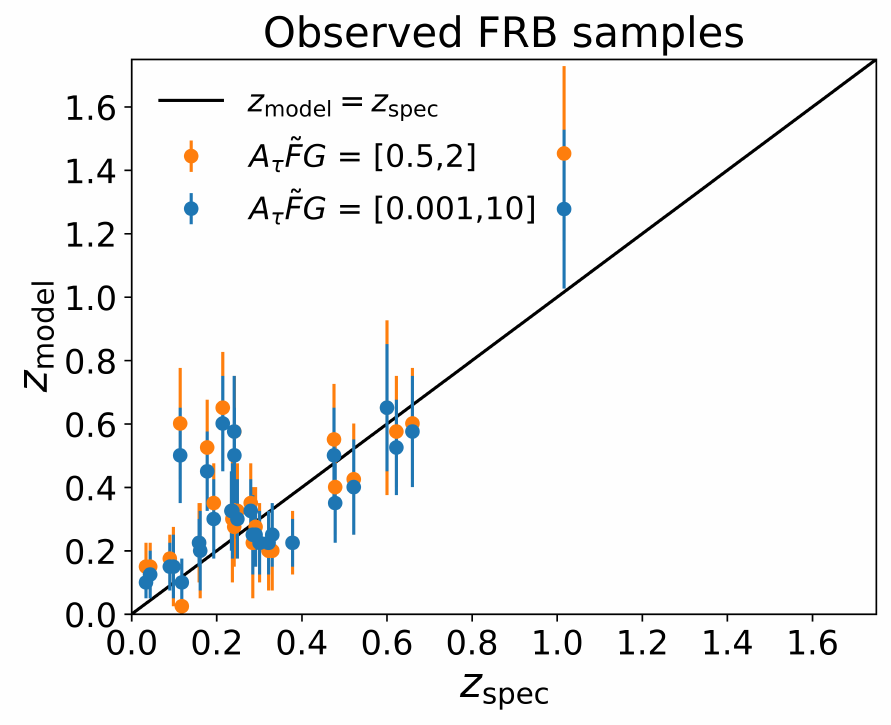}
        \includegraphics[width=\columnwidth]{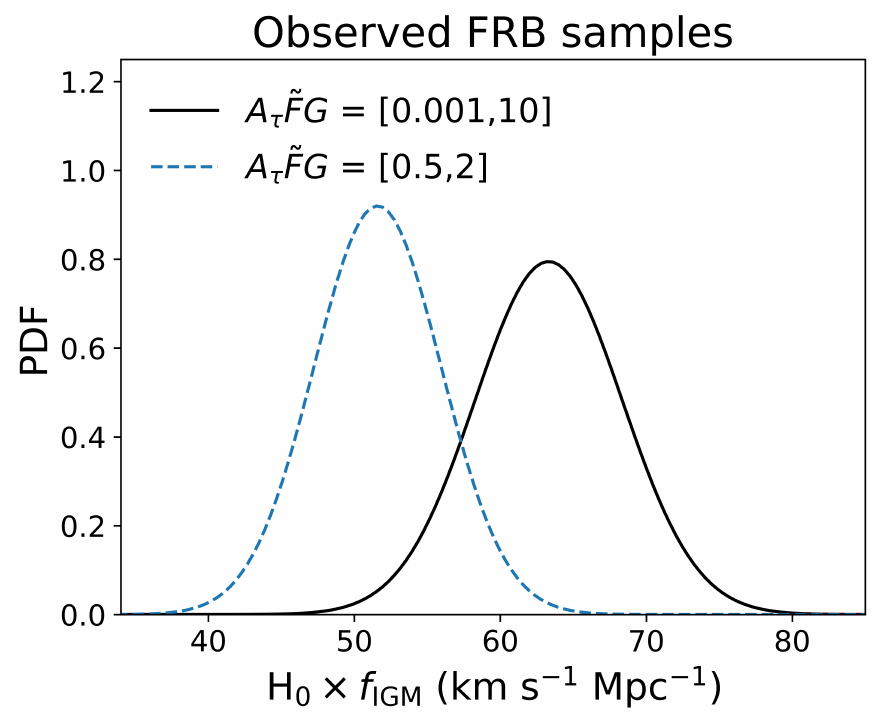}
        \includegraphics[width=\columnwidth]{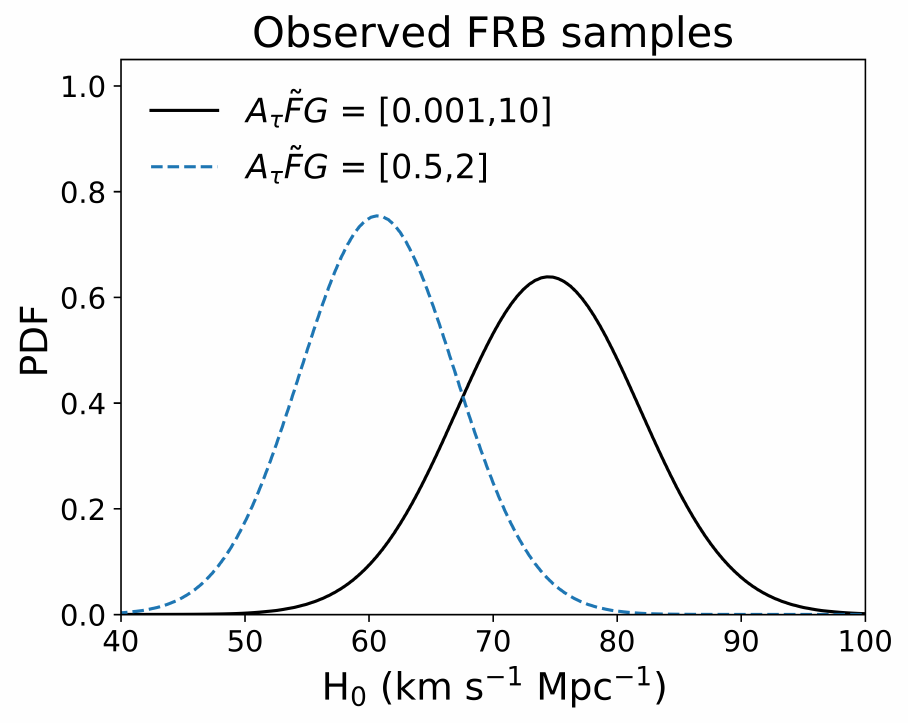}    
        \caption {Caparison between the wide and narrow ranges of AFG. Top panel: 
        Same as Fig. \ref{878}, but with optimized $f_{IGM}\times {\rm H}_{\rm 0}$ values, where $f_{\rm IGM}\times {\rm H}_{\rm 0}$ are $51.5^{+4.4}_{-4.4}$ km s$^{-1}$ Mpc$^{-1}$ and $63.2^{+4.8}_{-5.2}$ km s$^{-1}$ Mpc$^{-1}$ for the narrow and wide $A_{\tau} \tilde{F} G$ ranges, respectively.
        Middle panel: PDF of $f_{IGM}\times {\rm H}_{\rm 0}$ using the 30 FRB samples and our method.
        The solid black line corresponds to the wide $A_{\tau} \tilde{F} G$ range, and the dashed blue line corresponds to the narrow $A_{\tau} \tilde{F} G$ range. 
        Bottom panel: PDF of ${\rm H}_{\rm 0}$ for a given $f_{\rm IGM}=0.85\pm0.05$ using the 30 FRB samples and our method.
        }
        \label{87}
    \end{figure}

    We compared results based on the observed data, assuming narrow and wide ranges of $A_{\tau} \tilde{F} G =[0.5,2]$ and $[0.001,10]$ $({\rm pc}^2 \ {\rm km})^{-\frac{1}{3}}$, respectively. 
    Figure \ref{878} shows a comparison between the narrow and wide ranges of $A_{\tau} \tilde{F} G$ in $z_{\rm model}$ versus $z_{\rm spec}$.
    The figure includes our 30 FRB samples, where we fixed the value of $f_{\rm IGM}\times {\rm H}_{\rm 0}=63.2$ km s$^{-1}$ Mpc$^{-1}$ to highlight the difference between the $A_{\tau} \tilde{F} G$ ranges. 
    $\chi^{2}$ is 75.1 for the narrow range, and 37.9 is for the wide range. 
    We found that $z_{\rm model}$ with $A_{\tau} \tilde{F} G =[0.5,2]$ $({\rm pc}^2 \ {\rm km})^{-\frac{1}{3}}$ are systematically lower than that with $A_{\tau} \tilde{F} G =[0.001,10]$ $({\rm pc}^2 \ {\rm km})^{-\frac{1}{3}}$, indicating that the different prior assumptions on $A_{\tau} \tilde{F} G$ systematically affect the H$_{0}$ measurement.
    To demonstrate this point, we present the best-fit $z_{\rm model}$ to $z_{\rm spec}$ for two cases of the $A_{\tau} \tilde{F} G$ ranges by optimizing $f_{\rm IGM}\times {\rm H}_{\rm 0}$ in the top panel of Fig. \ref{87}. 
    The best-fit $f_{\rm IGM}\times {\rm H}_{\rm 0}$ (and $\chi^{2}$) are $51.5^{+4.4}_{-4.4}$ km s$^{-1}$ Mpc$^{-1}$ (58) and $63.2^{+4.8}_{-5.2}$ km s$^{-1}$ Mpc$^{-1}$ (37.9) for the narrow and wide ranges, respectively (middle panel of Fig. \ref{87}).
    Given a fixed value of $f_{\rm IGM}=0.85\pm0.05$, these correspond to ${\rm H}_{\rm 0}=60.6^{+6.3}_{-6.3}$ km s$^{-1}$ Mpc$^{-1}$ and ${\rm H}_{\rm 0}=74.3^{+7.2}_{-7.5}$ km s$^{-1}$ Mpc$^{-1}$ for the narrow and wide $A_{\tau} \tilde{F} G$ ranges, respectively (bottom panel of Fig. \ref{87})

    The $\chi^{2}$ value is higher for the narrow $A_{\tau} \tilde{F} G$ range, suggesting a worse fit to the observed data with $A_{\tau} \tilde{F} G =[0.5,2]$ $({\rm pc}^2 \ {\rm km})^{-\frac{1}{3}}$.
    This might suggest that the narrow range does not fully cover the true $A_{\tau} \tilde{F} G$ distribution in the FRB samples. The narrow-range model therefore does not fit the observed data well. 
    In contrast, the wide $A_{\tau} \tilde{F} G$ range is expected to perform better because its coverage in the $A_{\tau} \tilde{F} G$ parameter space is better.
    We speculate that the wide $A_{\tau} \tilde{F} G$ range covers the true $A_{\tau} \tilde{F} G$ distribution better, and it would be closer to the ideal situation of our simulation without significant systematics (Sect. \ref{priorAFG}) than the narrow $A_{\tau} \tilde{F} G$ range.
    In this sense, the wide $A_{\tau} \tilde{F} G$ range would be preferable in our analysis.
    
    Detailed investigations of the $A_{\tau} \tilde{F} G$ impact on $f_{\rm igm}\times {\rm H}_{\rm 0}$ have to be made and the optimal range of $A_{\tau} \tilde{F} G$ must be determined before our method can address the Hubble tension properly.
    Physically motivated constraints of the $A_{\tau} \tilde{F} G$ parameter would also play an important role in searching for the optimal $A_{\tau} \tilde{F} G$ range.
    We leave these points as future work because we focused on proposing a method for constraining H$_{0}$ with the scattering rather than emphasizing the current accuracy in this paper.

    \subsection{Future FRB samples}
    Currently, the number of localized FRBs is limited to $\sim40$ \citep[e.g.,][]{Bhandari_2022,Law2023}. 
    About 100 localized FRBs would be sufficient to clarify the Hubble tension with an uncertainty of $\sim$2.5 km s$^{-1}$ Mpc$^{-1}$ \citep{James_2022}. 
    More FRBs will be localized with scattering measurements with future instruments, including the CHIME outrigger \citep{Mena_Parra+2022}, the Deep Synoptic Array-110, ASKAP, and the Bustling Universe Radio Survey Telescope in Taiwan (BURSTT) \citep{2022PASP..134i4106L,2023ApJ...950...53H}.
    BURSTT can localize $\sim$100 FRBs per year \citep{2022PASP..134i4106L} to identify host galaxies with scattering measurements.
    Unlocalized FRBs can be used to derive the redshift statistically to further increase the samples \citep[e.g.][]{2022arXiv221213433Z}. 
    Therefore, our method can be tested with better statistics in the near future.
    
\section{Conclusions}

    A significant difference of 4 to 6 sigma exists in determining the Hubble constant (H$_0$) for two distinct methods, the cosmic microwave background (CMB) and the local distance ladders. This difference is most likely caused by unknown systematic errors. Therefore, devising an independent method for measuring ${\rm H}_{0}$ is the most important mission for addressing this unresolved puzzle.
    The FRBs offer a unique observable, DM$_{\rm IGM}$, which is a new distance indicator to derive ${\rm H}_{\rm 0}$. We summarize the result of this work below.
        \begin{enumerate}
      \item ${\rm DM_h}$ had to be assumed in previous works to derive DM$_{\rm IGM}$. 
        The scattering enabled us to measure ${\rm DM_{IGM}}$ with a parameter that combined a pulse profile, density fluctuations, and the geometry of a scattering screen ($A_{\tau} \tilde{F} G$ parameter).
        We used this parameterization for 30 FRBs with scattering measurements to model the redshifts, and we compared them with the observed redshifts (spectroscopic redshifts) to constrain H$_{0}$.
      
      \item We demonstrated that our method reduces the systematic error of H$_0$ by 9.1$\%$ compared to the previous method, and the statistical error is reduced by 1$\%$. 
      The reduction in systematic error is comparable to the Hubble tension ($\sim10\%$), indicating that our method can address the Hubble tension using future FRB samples.
      \item We measured a Hubble constant of ${\rm H}_{\rm 0}=74.3 ^{+7.2} _{-7.5}$  km s$^{-1}$ Mpc$^{-1}$ with our method using scattering. 
        The central value of this result prefers the measurement from the local distance ladder \citep[H$_{0}= 73.0 \pm 1.0$ km s$^{-1}$ Mpc$^{-1}$;][]{Riess_2022} over the CMB \citep[H$_{0}= 67.7 \pm 0.4$ km s$^{-1}$ Mpc$^{-1}$;][]{refId0}. 
        However, our measurement is still consistent with the two methods within an error of 1 $\sigma$.
      
   \end{enumerate}
    We described two future directions to reduce the error in our measurement. The first direction is to use more localized FRBs with scattering measurements, which are to be detected with future instruments, including BURSTT \citep{2022PASP..134i4106L}. BURSTT is the Taiwanese radio array, which can localize $\sim$100 FRBs per year to identify host galaxies with scattering \citep{2022PASP..134i4106L}. The second direction is to statistically treat unlocalized FRBs to further increase the samples \citep[e.g.][]{2022arXiv221213433Z}.

    \begin{acknowledgements}
    
    We would like to express our deepest appreciation to the anonymous referee for the comprehensive and thoughtful review of our manuscript. Their detailed examination and insightful suggestions have played a crucial role in refining our work, and the constructive feedback has greatly enhanced the overall quality and clarity of the paper.
    T-CY is grateful to Ms. Poya Wang for insightful discussions. 
    T-CY is also grateful to Dr. Shotaro Yamasaki for insightful discussions. 
    TG acknowledges the support of the National Science and Technology Council of Taiwan through grants 108-2628-M-007-004-MY3, 111-2112-M-007-021, and 112-2123-M-001-004-.
    TH acknowledges the support of the National Science and Technology Council of Taiwan through grants 110-2112-M-005-013-MY3, 110-2112-M-007-034-, 113-2112-M-005-009-MY3, and 113-2123-M-001-008-.
    We acknowledge the use of the CHIME/FRB Public Database, provided at \url{https://www.chime-frb.ca/} by the CHIME/FRB Collaboration.
    This research made use of Astropy, a community-developed core Python package for Astronomy \citep{Astropy2018}. 
    \end{acknowledgements}

%

%

\end{document}